# Using Agency Law to Tame AI Agents

Deven R. Desai* and Mark O. Riedl**

ABSTRACT

*Artificial Intelligence (AI) Agents have entered the marketplace. By some estimates, AI Agents, which promise to execute tasks and even take over your computer to do so, present a more than $400 billion market. The potential power of AI Agents has fueled legal scholars' fears that AI Agents will enable rogue commerce, human manipulation, rampant defamation, and infringement of intellectual property rights. These scholars are calling for regulation before AI Agents cause havoc.*

*This Article addresses concerns about AI Agents head-on. The Article leverages the socio-legal issues presented by agency law to identify potential problems. It then asks how computer science can solve those problems rather than trying to take rules for humans and impose them on software. This approach shows that core aspects of modern software engineering, particularly client-server architectures, address several of those concerns in ways quite different from—and arguably stronger than—how the law would discipline a human agent.*

* Sue and John Stanton Professor of Business Law and Ethics, Georgia Institute of Technology; J.D. Yale Law School; Affiliated Fellow, Yale Law School Information Society Project; former Academic Research Counsel, Google, Inc., Associate Director of Law, Policy, and Ethics, Georgia Tech Machine Learning Center. This Article was supported in part by summer research funding from the Scheller College of Business. We are grateful for the feedback from Professor Deborah Demott, especially on the law of agency, and from participants at the Legal Scholars Roundtable on Artificial Intelligence hosted by Emory Law School and at The Governance of Emerging Technologies and Science Conference hosted by the Sandra Day O'Connor College of Law, Beus Center for Law and Society, and Arizona State University. The views expressed herein are those of the authors alone and do not necessarily reflect the view of those who helped with and supported this work. All errors are ours.
** Professor, Georgia Tech School of Interactive Computing, Georgia Institute of Technology; Associate Director, Georgia Tech Machine Learning Center.

*The Article goes further and probes value alignment—a key part of current computer science best practices for mitigating potential agent harms—to show that agency law demands an update to value-alignment. Last, the Article draws on the theory of explainable AI to offer not only ways to address AI Agent's mistakes ex post, but also a novel ex ante option. Together, these changes enhance a user's ability to take action to correct or prevent AI Agent operations. These solutions will enable desired economic outcomes and mitigate perceived risks. In short, despite AI Agents' increasing capabilities, existing technical and socio-legal governance coupled with continual improvements in value alignment can enable them to meet the demands of law and society.*

TABLE OF CONTENTS

# I. INTRODUCTION

*For each possible percept sequence, a rational agent should select an action that is expected to maximize its performance measure, given the evidence provided by the percept sequence and whatever built-in knowledge the agent has.*

—Stuart Russell and Peter Norvig[1]

*The common law of agency, encompasses the legal consequences of consensual relationships in which one person (the "principal") manifests assent that another person (the "agent") shall, subject to the principal's right of control, have power to affect the principal's legal relations through the agent's acts and on the principal's behalf. Relationships of agency usually contemplate three parties—the agent, the principal, and third parties with whom the agent interacts in some manner.*

—Restatement (Third) of Agency[2]

AI Agents—software that perceives the world, has reasoning capabilities, and can act autonomously to carry out a user's instructions[3]—are out of the lab, in ever-growing use, and raising fears about runaway harms.[4] Perhaps because computer science uses the phrase "software agents" and certain sectors of the software industry talk of "agentic software," legal discussions have called for society "to take seriously the word agent in its legal and social sense" and treat software agents as human agents.[5] This Article takes this call head-on.

---

1. STUART J. RUSSELL & PETER NORVIG, ARTIFICIAL INTELLIGENCE: A MODERN APPROACH 40 (4th ed. 2021).

2. RESTATEMENT (THIRD) OF AGENCY INTRODUCTION (A.L.I. 2006).

3. RUSSELL & NORVIG, *supra* note 1, at 3–4 ("[A]ll computer programs do something, but computer agents are expected to do more: operate autonomously, perceive their environment, persist over a prolonged time period, adapt to change, and create and pursue goals."); *cf.* Maxwell Zeff, *AI Agents Promise to Connect the Dots Between Reality and Sci-Fi*, GIZMODO (May 27, 2024), https://gizmodo.com/ai-agents-openai-chatgpt-google-gemini-reality-sci-fi-1851500474 ("Simply put, AI Agents are just AI models that do something independently . . . They go a step further than just creating a response like the chatbots we've become familiar with—there's action . . . . [Companies are] teaching AI Agents to work with various APIs on your computer. Ideally, they can press buttons, make decisions, autonomously monitor channels, and send requests.").

4. Jonathan Zittrain, *We Need to Control AI Agents Now*, ATLANTIC (July 2, 2024), https://www.theatlantic.com/technology/archive/2024/07/ai-agents-safety-risks/678864/ (describing AI Agents perpetrating broad, harmful acts online). Professors Ian Ayres and Jack Balkin argue AI Agents should be treated as unmanaged, risky agents. Ian Ayres & Jack Balkin, *The Law of AI Is the Law of Risky Agents Without Intentions*, U. CHI. L. REV. ONLINE, Nov. 2024, at 6–8, https://lawreview.uchicago.edu/sites/default/files/2024-11/Ayres_Balkin_Law%20of%20Risky%20Agents.pdf (discussing fears of defamatory AI Agents).

5. Jonathan Zittrain, Kevin Frazier & Jen Patja, *Lawfare Daily: Jonathan Zittrain on Controlling AI Agents*, LAWFARE (Oct. 17, 2024), https://www.lawfaremedia.org/article/lawfare-daily--jonathan-zittrain-on-controlling-ai-agents (positing "even ordering Domino's via an agent could lead

We argue that agency law provides insights about potential problems AI Agents raise, but that direct application of agency law to AI Agents is not the best way to understand AI Agents, let alone manage them.[6] Instead, we treat the law of agency and how to manage human agents as a specification of harms to avoid and goals to meet. That premise allows us to identify potential concerns and then offer solutions that map to computer science realities. AI Agents raise classic law of agency concerns such as setting authority, preventing agents from taking rogue actions, and third-party knowledge issues. But, despite provocative rhetoric around similarity to human brains or human workers, AI Agents are fundamentally software. As such, rather than trying to make AI Agents operate through incentives and penalties like a human agent, one part of the solution draws on APIs—a core piece of the internet's technical infrastructure and especially ecommerce—to address concerns around authority and preventing misbehavior.[7] In addition to APIs, we draw on the growing AI software engineering practice of value-alignment training to show how that practice addresses concerns about AI Agents becoming tools for bad actors.[8] We then analyze the law of agency and computer science's theory of agents and find that the legal issues concerning fiduciary duties of loyalty and conflict of interest are missing in current computer science. This Article argues that these fiduciary duties should be used as part of computer science's toolkit for alignment.[9] Together, these approaches address socio-legal issues around AI Agents and offer a path to building responsible AI Agents.[10]

Imagine prompting ChatGPT, Alexa, Gemini, or a software system that you built with the following: "Plan *and book* a two-week trip to Bora Bora. I want some beach time but also some cool experiences." Within an hour, your phone

---

to unintended and severe consequences" such ordering "a thousand pizzas"); *see also* SIMON GOLDSTEIN & PETER N. SALIB, AI RIGHTS FOR ECONOMIC FLOURISHING (Dec. 15, 2025), https://papers.ssrn.com/sol3/papers.cfm?abstract_id=5353214 (arguing that artificial general intelligence will soon be here and then such software should be treated as humans).

6. Indeed, importing agency law's view of agents as representative of humans arguably plays into the hands of those who wish to argue that software has or will achieve Artificial General Intelligence and even consciousness.

7. *See infra* Section D. (identifying where the law of agency and the computer science view of agents converge and diverge on issues around agents).

8. *See infra* Section IC.

9. *See infra* notes 31–34 and accompanying text. For an early take on the connection between the duty of loyalty and software design, see Anthony Aguirre, Gaia Dempsey, Harry Surden & Peter B. Reiner, *AI Loyalty: A New Paradigm for Aligning Stakeholder Interests*, 1 IEEE TRANSACTIONS ON TECH. & SOC'Y 128 (2020).

10. The term "responsible" begs the question: responsible to whom? This Article explores responsibility from the perspectives of communities, which might be society as a whole or a marketplace of principles, suppliers, and third parties, and also individual users. The concept of value alignment, while most commonly invoked in the context of constraining AGI from hypothetical harms to humanity writ large, is also pragmatically a tool for ensuring that AI systems are more capable of working for the best interests of individual users. *See infra* Section IC. We also invoke the concept of explanations so that agents provide users opportunities to better understand when and how to reverse agent actions. *See infra* Section B.

pings, and you see an amazing trip that you love, fully booked and paid for.[11] Now imagine you operate a business and have a new product but are quite busy.[12] You tell the software, "Write a series of social media posts based on these materials (which you add to the prompt). Have each post highlight the novel features. Deploy the posts across X, Facebook, TikTok, YouTube, Instagram, Snapchat, and Reddit for the next two weeks in ways appropriate for each platform." Over the next few weeks, the posts go out, and your sales go up. Your experience went so well you use the software to handle sales. This time, however, the software makes a mistake and sells your product for half its value.[13] Now consider a situation where someone hostile to the U.S. government instructs the software to generate 100,000 social media posts falsely claiming that Donald Trump is a paid puppet of Russia and another 100,000 posts falsely claiming that Nancy Pelosi relies on DeepSeek's AI because she is a paid operative for China. The person directs the software to make the posts appear to come from diverse individuals and to deploy them over two weeks to simulate growing public support.[14] These hypotheticals *seem* possible because of the advent of AI Agents. Moreover, industry has embraced and is offering AI Agent software as the next big wave of services flowing from the investment in generative AI.[15] Thus, legal scholars have begun to question the implications of such software.

---

11. *See* Clay Bavor & Bret Taylor, *The Guide to AI Agents*, SIERRA.AI (May 23, 2024), https://sierra.ai/news/ai-agents-guide ("[AI] agents are autonomous software systems that can reason, make decisions, and pursue goals with creativity and flexibility, all while staying within the bounds that have been set for them. Whereas applications *help* you do the work, agents get the work done for you."); Michelle Castillo, *Warren Buffett Says One Question Posed by AI Has Stumped Economists for a Century*, CNBC (May 7, 2024), https://www.cnbc.com/2024/05/07/warren-buffett-on-ai-issue-that-has-stumped-economists-for-a-century.html ("Clearly, there is a trend where we go to 'agentic' workflows, agents take actions on behalf of the end user autonomously. It's a little ways out, but people do need to build apps and enrich customer experience and drive more cost out of the business and find new ways to drive growth."); *see* Alex Cosmas & Vik Krishnan, *What AI Means for Travel—Now and in the Future*, MCKINSEY & CO. (Nov. 2023), https://www.mckinsey.com/industries/travel-logistics-and-infrastructure/our-insights/what-ai-means-for-travel-now-and-in-the-future#/ (noting generative AI may save trillions across sectors).

12. While writing this article, a real-world example emerged that closely resembles the given hypothetical. *See* Erin Griffith, *How A.I. Helped One Man (and His Brother) Build a $1.8 Billion Company*, N.Y. TIMES (Apr. 2, 2026), https://www.nytimes.com/2026/04/02/technology/ai-billion-dollar-company-medvi.html. The company CEO used AI tools to set up a commerce site, generate ads, and handle customer service. *Id.* When the software erred on prices, he honored what customers were quoted, and he continually tested the software to prevent additional errors. *Id.*

13. A similar event recently happened when Air Canada used software for customer service, and the software promised a price that the airline did not want to offer. *See* Maria Yagoda, *Airline Held Liable for its Chatbot Giving Passenger Bad Advice - What This Means for Travellers*, BBC NEWS (Feb. 23, 2024), https://www.bbc.com/travel/article/20240222-air-canada-chatbot-misinformation-what-travellers-should-know; *see also* Billy Perrigo, *Exclusive: Anthropic Let Claude Run Its Office Shop. Then Things Got Weird*, TIME (June 27, 2025), https://time.com/7298088/claude-anthropic-shop-ai-jobs/ (noting the system was convinced to sell items at a loss among other oddities); *Project Vend: Can Claude Run a Small Shop? (And Why Does That Matter?)*, ANTHROPIC (June 27, 2025), https://www.anthropic.com/research/project-vend-1 (documenting what worked and what failed in the company's experiment with letting an AI Agent manage an experimental small shop); Michael Nuñez, *Can AI Run a Physical Shop? Anthropic's Claude Tried and The Results Were Gloriously, Hilariously Bad*, VENTURE BEAT (June 27, 2025), https://venturebeat.com/ai/can-ai-run-a-physical-shop-anthropics-claude-tried-and-the-results-were-gloriously-hilariously-bad/ (noting the system turned down an offer that was 567% markup and began hoarding tungsten cubes for a grab-and-go food and drink fridge).

Whether AI Agents will be commercially significant is not a purely academic question.[16] Generative AI services have yet to offer clear economic payoffs.[17] Industry proponents assert that AI Agents present a way to realize more immediate returns on the multi-billion-dollar investment in generative AI to date.[18] Companies such as OpenAI, Google, Microsoft, and Amazon[19] hope that AI Agents—as offshoots of the computer science breakthroughs behind generative AI—will autonomously take over all sorts of tasks currently performed by humans.[20] If so, AI Agents will aid individuals and industries as a cost-reducing, productivity-enhancing technology.[21]

---

14. *Cf.* Ayres & Balkin, *supra* note 4, at 6–8 (discussing fears of defamatory AI Agents); Zittrain, *supra* note 4 (describing AI Agents perpetrating broad, harmful acts online).

15. *See infra* notes 17–21 and accompanying text.

16. *See* Cade Metz & Nico Grant, *Google Unveils A.I. Agent That Can Use Websites on Its Own*, N.Y. TIMES (Dec. 11, 2024), https://www.nytimes.com/2024/12/11/technology/google-ai-agent-gemini.html; Maxwell Zeff, *The Race Is on to Make AI Agents Do Your Shopping for You*, TECHCRUNCH (Dec. 2, 2024), https://techcrunch.com/2024/12/02/the-race-is-on-to-make-ai-agents-do-your-online-shopping-for-you/ (noting Perplexity's release of its shopping AI Agent); Mike Bird & Alice Fulwood, *Special (Offers) Agent: Why Retailers Should Be Worried about AI.*, MONEY TALKS, ECONOMIST (Nov. 27, 2025), https://www.economist.com/podcasts/2025/11/27/special-offers-agent-why-retailers-should-be-worried-about-ai (noting a McKinsey Consulting claim that "close to 55%" of consumers are using generative AI to shop). At a broader level, AI Agents raises issues around of the law of agency and its practical implications. *See* Deborah A. DeMott, *Disloyal Agents*, 58 ALA. L. REV. 1049, 1067 (2007) ("The common law of agency has not always attracted the degree of academic interest that's warranted by its ubiquity, as well as its theoretical interest and practical significance.").

17. *See* Megan Cerullo, Here's Who Is Spending Money on AI Subscriptions, and How Much They Cost, CBS NEWS, (April 17, 2026), https://www.cbsnews.com/news/generative-ai-subscriptions-consumer-spending/ (noting only 2% of households pay for GenAI subscriptions as compared to 25% paying for streaming or 5% spending on sports betting apps and the challenge to change free users to paying users); Danielle Kost. Most GenAI Players Remain "Far Away" from Profiting: Interview with Andy Wu, HARVARD BUSINESS SCHOOL, WORKING KNOWLEDGE (Oct. 30, 2025), https://www.library.hbs.edu/working-knowledge/many-gen-ai-players-remain-far-away-from-profiting-interview-with-andy-wu, (explaining the returns on investment "remains murky" with subscription costs "unlikely to cover the costs"); *see also* Castillo, *supra* note 11 (noting difficulty in parsing older AI software deployment from generative AI when assessing productivity returns); Scott Rosenberg, *Generative AI Is Still a Solution in Search of a Problem*, AXIOS (Apr. 24, 2024), https://www.axios.com/2024/04/24/generative-ai-why-future-uses ("The gigantic and costly industry Silicon Valley is building around generative AI is still struggling to explain the technology's utility."). *But see* Vala Afshar, *How AI Agents Can Generate $450 Billion by 2028 - and What Stands in the Way*, ZD NET (July 21, 2025), https://www.zdnet.com/article/how-ai-agents-can-generate-450-billion-by-2028-and-what-stands-in-the-way/ (discussing CapGemini Research Institute report on the market for AI Agents).

18. *See* Cristina Criddle & George Hammond, *OpenAI Bets on AI Agents Becoming Mainstream by 2025*, FIN. TIMES (Oct. 1, 2024), https://www.ft.com/content/30677465-33bb-4f74-a8e6-239980091f7a; James O'Donnell, *Sam Altman Says Helpful Agents Are Poised to Become AI's Killer Function*, MIT TECH. REV. (May 1, 2024), https://www.technologyreview.com/2024/05/01/1091979/sam-altman-says-helpful-agents-are-poised-to-become-ais-killer-function/; Castillo, *supra* note 11 ("The market is passing through the phase of the value accruing only at the bottom layer, such as Nvidia and ChatGPT/OpenAI, and it is now critical for companies to prepare for the applications built on top of that infrastructure."); Bavor & Taylor, *supra* note 11 ("Behind this breakthrough [in AI Agents] are recent advances in artificial intelligence, and large language models in particular").

19. *See infra* notes 53–57 and accompanying text.

At the same time, the proliferation of companies and individuals using AI Agents raises age-old issues for commerce conducted at a distance, including: what constitutes a binding contract, how payment is guaranteed, and how erroneous purchases should be handled.[22] As AI Agents grow in use and power, some are concerned about monitoring costs and the difficulty for humans to overcome information asymmetry.[23] There are also claims that AI Agents may take "highly unpredictable or unintuitive actions."[24] For example, one fearful perspective is that AI Agents will enable "unintended and severe consequences" such as massive Domino pizza orders[25] or wild, million-dollar purchasing errors.[26] Even if such extremes are unlikely, a simpler concern exists. Human agents can, after all, err in carrying out mundane but important commercial tasks too. Given AI Agents' potential speed, they might err at a greater rate and cost than human agents. Although the issues around commerce are important, online life presents other avenues for undesired outcomes.

Another fear is that AI Agents may enable rogue, harmful actions unless we do something to "control" them immediately.[27] For example, a prominent scholar posits that an AI Agent taking a simple instruction such as, "help me cope with this boring class" could call in a bomb threat to carry out the request, or an army of AI Agents might manipulate people to make bomb threats at scale.[28] Additional concerns include AI Agents that could lurk in an online game and join online fan forums but with the ulterior motive of slipping in a political

20. YONADAV SHAVIT, SANDHINI AGARWAL, MILES BRUNDAGE, STEVEN ADLER, CULLEN O'KEEFE, ROSIE CAMPBELL, TEDDY LEE, PAMELA MISHKIN, TYNA ELOUNDOU, ALAN HICKEY, KATARINA SLAMA, LAMA AHMAD, PAUL MCMILLAN, ALEX BEUTEL, ALEXANDRE PASSOS & DAVID G. ROBINSON, PRACTICES FOR GOVERNING AGENTIC AI SYSTEMS 6–7 (Dec. 2023), https://cdn.openai.com/papers/practices-for-governing-agentic-ai-systems.pdf (describing a range of "potential benefits" of AI Agent systems).

21. Castillo, *supra* note 11; *cf.* Cosmas & Krishnan, *supra* note 11 (noting generative AI may save trillions across sectors).

22. *See, e.g.*, Geoffrey Fowler, *I Let ChatGPT's New 'Agent' Manage My Life. It Spent $31 on a Dozen Eggs*, WASH. POST (Feb. 7, 2025), https://www.washingtonpost.com/technology/2025/02/07/openai-operator-ai-agent-chatgpt/.

23. Noam Kolt, *Governing AI Agents*, 101 NOTRE DAME L. REV. (forthcoming) (manuscript at 33–34) (comparing human agency to software agent issues), https://papers.ssrn.com/sol3/papers.cfm?abstract_id=4772956. As discussed below, software offers the potential for better monitoring and solutions to information asymmetry than possible in a purely human to human context.

24. *Id.* at 34.

25. *See* Zittrain, *supra* note 5 (positing that "even ordering Domino's via an agent could lead to unintended and severe consequences" such as ordering "a thousand pizzas").

26. *See* Zittrain, *supra* note 5; Zittrain, *supra* note 4.

27. Zittrain, *supra* note 4. Kolt offers a list of concerns including financial loss, property damage, threats to life, and autonomous hacking of websites. Kolt, *supra* note 23, at 16. These examples, however, are not clear as to whether principal-agent concerns are at stake or the more general fear of "powerful technology" is the issue.

28. Zittrain, *supra* note 4.

agenda.[29] Other scholars envision AI Agents as "risky agents" and seek to discipline the way the software is created so as to minimize harmful outcomes, such as copyright infringement and defamation potentially flowing from using AI Agents.[30]

We argue that most of the concerns around AI Agents revolve around solving how best to ensure that a piece of software does what a user wants the software to do, or what computer science broadly calls "the alignment" problem.[31] Alignment is about how well software performs given a specification by the programmer of the software. For example, when writing software that allows a robot vacuum to clean a room, one likely will add code so that the robot does not damage furniture or spread dirt around the house.[32] Of late, the alignment problem, broadly speaking, often addresses more abstract goals, such as avoiding biased outputs and notions of "doing the right thing."[33] No matter the goal or the underlying technology driving the AI—including generative AI and robotics—the more dynamic the operating environment and the more a software can learn, the less one can fully specify what one wants the software agent to do.[34] Thus, alignment becomes quite difficult. This point links to legal concepts.

The gaps in how well we can specify software behavior track several core issues in agency law.[35] Both computer science and agency law are concerned with how to have an agent do something for someone, how to specify instructions for an agent, how to have an agent perform a task with imperfect instructions, and how to limit the agent's deviation from the instructions. Agency law, however, goes a step further and asks about the way agents affect third parties.[36]

29. *Id.*

30. Ayres & Balkin, *supra* note 4, at 6–11.

31. *See infra* Section IC.

32. *See, e.g.*, Brain Heater, *iRobot's Poop Problem*, TECHCRUNCH (Sept. 9, 2021), https://techcrunch.com/2021/09/09/actuator-4/ (detailing extensive research by iRobot on how to have its vacuum not pick up and smear dog feces around a house); Samantha Nelson, *'Pooptastrophe': Man Details the Night His Roomba Ran Over Dog Poop*, USA TODAY (Aug. 15, 2016), https://www.usatoday.com/story/news/nation-now/2016/08/15/pooptastrophe-man-details-night-his-roomba-ran-over-dog-poop/88667704/.

33. RUSSEL & NORVIG, *supra* note 1, at 57. As an example of a failure to align, in 2016 when Microsoft released its chatbot, Tay, designed to interact on Twitter, they didn't want the software to give racist outputs; but user interaction yielded that undesired result. *See* Marty J. Wolf, K. Miller & Frances S. Grodzinsky, *Why We Should Have Seen That Coming: Comments on Microsoft's Tay "Experiment," and Wider Implications*, 47 ACM SIGCAS COMPUTERS & SOC'Y 54, 54–55 (2017).

34. *See infra* notes 77–86 and accompanying text.

35. *See infra* Section C.; *accord* Dylan Hadfield-Menell & Gillian K. Hadfield, *Incomplete Contracting and AI Alignment*, *in* PROCS. 2019 AAAI/ACM CONF. ON AI, ETHICS, & SOC'Y 417, 417 (2019) (connecting alignment issues in computer science to principal-agent issues in economics and law).

36. Kolt's work offers that "[a]t its core, agency law" is about the relationship between principals and their agents and focuses on harms to principals. *See supra* note 23, at 20–21. The work does not address third-party issues regarding trust and the way agency law is also about the vital need to protect third parties in the contracts context.

This issue remains underdeveloped, if not overlooked, in legal scholarship and is not explicitly addressed by computer science's theory of agents.

Even though computer science's theory of agents has a narrow view of interaction with users that does not account for third parties, computer science has solutions to core issues raised by agency law, such as monitoring, setting authority boundaries, and limiting the possibility of undesired actions.[37] Indeed, core parts of why software works and the way AI Agents interact with other software provide robust answers to concerns over potentially undesired outcomes when AI Agents are used.[38] We thus add to the literature on AI Agents by examining the broad set of technical infrastructure and rules that limit AI Agents and by extension address issues legal scholars have raised but not solved.

Although technical infrastructure addresses several agency-related issues, other issues persist. As such, we go beyond technical approaches and examine how the common law of agency, especially its understanding of conflicts of interest, provides rich insights that can and should inform future value alignment of AI Agents as a soft law approach to responsible AI development.[39] In short, technical infrastructure and enhanced value alignment can mitigate agency-like concerns without converting software into legal agents.[40] Put differently, we argue that approaches to governing AI Agents can benefit from looking at how the law governs human agency, but only by analogy.

It is a mistake to equate a software agent to a human agent.[41] Although computer science uses the word agent, and computer science's theory of software

37. Kolt's view of enforcement assumes that the common law ways "for governing agent behavior" are the ones to apply to AI Agents and they come up short. *See* Kolt, *supra* note 23, at 36. This view seems to forget Kolt's own point that "[legal and economic] frameworks, however, appear ill-suited to addressing agents that operate very differently to human beings." *Id.* at 33. In contrast, this paper uses the economic and legal issues to ask what computer science options address the issues in ways suitable for software.

38. In that sense, critics should remember that AI Agents will have to operate within external structures and interact with institutions. *Cf.* Hadfield-Menell & Hadfield *supra* note 35, at 417 ("human contracting is supported by substantial amounts of external structure, such as generally available institutions (culture, law)").

39. As a matter of legal theory and approaches to governance, using the common law of agency to inform value alignment fits in the "soft law" approach to governing AI systems, where soft law is described as "a program that creates substantive expectations, but which are not directly enforceable by government." *See* Gary Marchant, Lucille Tournas & Carlos Ignacio Gutierrez, *Governing Emerging Technologies Through Soft Law*, 61 JURIMETRICS J. 1, 5 (2020). Marchant, Tournas, and Gutierrez describe "eight common themes in [] soft law programs: accountability, fairness and nondiscrimination, human control of technology, privacy, professional responsibility, promotion of human values, safety and security, and transparency and explainability." *Id.* at 6. The common law of agency can be understood as identifying similar themes but with greater specification of what is or isn't desired. Thus, it arguably serves as a distinct statement about high level issues while also providing greater, actionable specifications for computer scientists to address as they build systems.

40. *See, e.g.*, SAMIR CHOPRA & LAURENCE F. WHITE, A LEGAL THEORY FOR AUTONOMOUS ARTIFICIAL AGENTS 69 (2011) (concluding that as agents become ever more independent, intentional law will need to embrace legal personhood for agents).

41. *See, e.g.*, Ayres & Balkin, *supra* note 4, at 3–4 (anthropomorphizing AI Agents by asking that software not be negligent, exercise fiduciary care, and presuming "AI programs intend the agents looks quite similar to legal conceptions of agency, software lacks agency.[42] Anthropomorphizing software confuses issues and could lead to a world where software has legal personhood, related rights, and liability shields.[43] If so, expanded reliance on software could erode accountability—a result to be avoided.[44]

This Article proceeds in three Parts. Part II explains what AI Agents are, the range of actions AI Agents can take, and the emerging market for AI Agents. It adds to the literature on AI and the law by comparing the ways law and computer science conceive of agents in ways that converge and diverge. This investigation reveals that although the disciplines share terminology and some theoretical concerns, important differences mean it is a mistake to assume that the agency law is a direct path to governing AI Agents.[45] Part III shows that fear surrounding rogue AI agents in commercial settings misses a key point. Ecommerce already addresses issues around verifying commercial transactions at a distance and mitigating bot actions. In short, core aspects of the way software interacts with other software currently enables more than $1 trillion in U.S. online retail[46] and will also manage AI Agents' place in that sector. Part IV

reasonable and foreseeable consequences of their actions."). To be clear, Ayres and Balkin end up admitting that software lack intentions and the real focus should be on "question of legal obligation is who should be held responsible for the use of AI and under what conditions." *Id.* at 1. The danger lies in how legal scholars implicitly or explicitly talk about software agents in terms that move a reader to ascribe actual human capabilities to software.

42. Just because software looks like and acts like an agent, doesn't make it an agent in the legal sense of the word. Yet legal scholars are following the "duck test" for AI Agents. Ayres and Balkin imply software is an agent as shown by the title of their piece and their view of intent. *Id.* at 3–4 ("the law should ascribe intentions to AI programs" including the "foreseeable" outcomes of their actions). Zittrain is more extreme and offers, "an agent is meant to represent a person" in ways that map to social and legal understandings. *See* Zittrain, *supra* note 5. Yet, even strong supporters of AI Agent software are careful not to collapse the difference between human abilities to set goals and systems that are fully autonomous. *See* SHAVIT ET AL., *supra* note 20, at 5 ("We emphasize that agenticness [sic] is a distinct concept from consciousness, moral patienthood [sic], or self-motivation, and distinguish a system's degree of agenticness [sic] from its anthropomorphism.").

43. *See infra* notes 61–64 and accompanying text discussing how Air Canada tried to avoid liability by "suggest[ing] the chatbot is a separate legal entity that is responsible for its own actions." Moffatt v. Air Canada, 2024 BCCRT 149, para. 27 (Can. B.C. Civ. Resol. Trib.), https://decisions.civilresolutionbc.ca/crt/crtd/en/525448/1/document.do.

44. *Cf.* Deven R. Desai, *The Chicago School Trap in Trademark: The Co-Evolution of Corporate, Antitrust, and Trademark Law*, 37 CARDOZO L. REV. 551 (2015) (tracing the history of business associations and showing the shift from bounded business entities to unlimited action by such entities).

45. *Accord*, Kolt, *supra* note 23, at 37–38 (concluding that direct application of agency law mechanisms to AI agents "do not have simple analogs for AI Agents"). In contrast to our approach, Kolt's solutions turn to broad notions of inclusivity, visibility, and liability that either do not appreciate current possible technical solutions to the issues or do not provide concrete specifications for technologists and in that sense do not solve the issues.

46. *See E-commerce as a Share of Total U.S. Retail Sales from 1st quarter 2010 to 3rd Quarter 2025*, STATISTA, https://www.statista.com/statistics/187439/share-of-e-commerce-sales-in-total-us-retail-sales-in-2010/ (on file with Berkeley Technology Law Journal).

synthesizes insights from law and computer science to offer a path towards building legally informed AI Agents. The Article then concludes.

## II.
## CONCEPTIONS OF AGENTS AND AGENCY

The possibilities of software doing something for us, and the realities that software may act in ways that pose problems, connect to a fundamental problem. We cannot do everything by ourselves; agents extend our reach and capacity. But here we must be careful. As in other intersections between law and computer science, shared terminology masks important differences in meaning and implication.[47] For some aspects of agency, the two areas ask similar questions and seek similar limits. Yet there are key differences between the two views. In computer science, agents are tools or instruments that augment our capacity.[48] In contrast, legal agency extends the principal's legal personality.[49] Put simply, although both law and computer science use the term "agent," the underlying understanding in either discipline is different. This Part describes the nature and range of AI Agent products and services to lay a foundation to understand what AI Agents can do and how they operate. This Part then investigates the similarities and differences between computer science and legal views of agency to identify the gaps between the areas.

### *A. AI Agents: The Next Wave of Generative AI Tools*

The hope that AI Agents will generate huge productivity gains is fueling rapid investment in and proliferation of AI Agent research and services.[50] AI Agents are often defined as "autonomous software systems that can reason, make decisions, and pursue goals with creativity and flexibility . . . for you."[51] AI

47. *See, e.g.*, Deven R. Desai & Joshua A. Kroll, *Trust But Verify: A Guide to Algorithms and the Law*, 31 HARV. J. L. & TECH 7–11 (2018) (parsing the meanings of transparency and accountability in law and computer science); *cf.* Desai, *supra* note 44 (showing that competition in trademark law and competition in antitrust and corporate law have different origins and meanings); *see infra* Section B.

48. Although corporate rhetoric around "agentic" AI can trick people into thinking software has conscious intent, but that is not so.

49. We thank Professor Demott for noting the distinction and encouraging us to draw it out.

50. *Cf.* Steven Rosenbush, *CEOs Are All-In on AI*, The Wall Street Journal, December 25, 2025, https://www.wsj.com/articles/ceos-are-all-in-on-ai-f3882564 (noting survey of more than 100 U.S. companies with more than 10,000 employees shows that CEOs believe AI will increase "productivity, help the economy, improve the global economy, improve competitiveness").

51. Bavor & Taylor, *supra* note 11; *accord* K.R. CHOWDHARY, FUNDAMENTALS OF ARTIFICIAL INTELLIGENCE 473 (2020) (listing autonomous, adaptable, knowledgeable, mobile, collaborative, persistent as characteristics of AI Agents); IASON GABRIEL, ARIANNA MANZINI, GEOFF KEELING, LISA ANNE HENDRICKS, VERENA RIESER, HASAN IQBAL, NENAD TOMAŠEV, IRA KTENA, ZACHARY KENTON, MIKEL RODRIGUEZ, SELIEM EL-SAYED, SASHA BROWN, CANFER AKBULUT, ANDREW TRASK, EDWARD HUGHES, A. STEVIE BERGMAN, RENEE SHELBY, NAHEMA MARCHAL, CONOR GRIFFIN, JUAN MATEOS-GARCIA, LAURA WEIDINGER, WINNIE STREET, BENJAMIN LANGE, ALEX INGERMAN, ALISON LENTZ, REED ENGER, ANDREW BARAKAT, VICTORIA KRAKOVNA, JOHN OLIVER SIY, ZEB KURTH-NELSON, AMANDA MCCROSKERY, VIJAY BOLINA, HARRY LAW, MURRAY SHANAHAN, LIZE ALBERTS, BORJA BALLE, SARAH DE HAAS, YETUNDE IBITOYE, ALLAN DAFOE,

Agents are "proactive," as they interact with other agents and adapt to changes in the world around the agent on behalf of the user.[52] And AI Agents are now out of the lab and rolling out at a rapid rate. For example, OpenAI has released its Operator AI Agent.[53] OpenAI frames the software as a step toward AI Agents that function as "super-competent colleague[s]," equipped with comprehensive user knowledge and autonomous task-execution capabilities.[54] Google and Microsoft now offer AI Agent services to navigate and take actions on the Web.[55] Amazon's AWS division touts AI Agents as "rational agents" that will provide improved productivity, reduced costs, informed decision-making, and improved customer experiences.[56]

What distinguishes these from prior Generative AI systems is that instead of simply talking about what you should do, AI Agents can do the things it suggests on your behalf. That is, rather than just giving you a detailed itinerary for a great week of things to do with children on a break from school, your AI Agent would plan and book day camps, craft fairs, and zoo trips, just as you might do after a series of web searches.[57]

Beyond executing tasks using existing services, AI Agents also promise to create new things. For example, instead of having a software team use generative

---

BETH GOLDBERG, SÉBASTIEN KRIER, ALEXANDER REESE, SIMS WITHERSPOON, WILL HAWKINS, MARIBETH RAUH, DON WALLACE, MATIJA FRANKLIN, JOSH A. GOLDSTEIN, JOEL LEHMAN, MICHAEL KLENK, SHANNON VALLOR, COURTNEY BILES, MEREDITH RINGEL MORRIS, HELEN KING, BLAISE AGÜERA Y ARCAS, WILLIAM ISAAC & JAMES MANYIKA, THE ETHICS OF ADVANCED AI ASSISTANTS 7 (Apr. 28, 2024), https://arxiv.org/abs/2404.16244 (defining "an AI assistant as an artificial agent with a natural language interface the function of which is to plan and execute sequences of actions on the user's behalf across one or more domains and in line with the user's expectations") .

52. CHOWDHARY, *supra* note 51, at 472 (listing "autonomous, adaptable, knowledgeable, mobile, collaborative, persistent" as characteristics of AI Agents).

53. Sunny Yadav, *OpenAI Agent 'Operator' to Handle Complex Tasks for People Starting 2025*, EWEEK (Nov. 29, 2024), https://www.eweek.com/news/openai-agent-handles-tasks-for-people/.

54. O'Donnell, *supra* note 18.

55. Metz & Grant, *supra* note 16; Abner Li, *Report: Google Preps 'Jarvis' AI Agent That Works in Chrome*, 9TO5GOOGLE (Oct. 26, 2024), https://9to5google.com/2024/10/26/google-jarvis-agent-chrome/; *see* Nigel Powell, *Microsoft Unveils Magentic-One—an AI Agent That Can Browse the Web and Write Code*, TOM'S GUIDE (Nov. 13, 2024), https://www.tomsguide.com/ai/microsoft-unveils-magentic-one-an-ai-agent-that-can-browse-the-web-and-write-code (Microsoft has launched Magentic-One, which uses an "Orchestrator" agent to manage other agents "a WebSurfer, FileSurfer, Coder and ComputerTerminal."). The hope is that you could instruct the "Orchestrator" and it could use the sub-agents as specialists to carry out different needed tasks. *Id.*

56. Amazon Web Services, *What Are AI Agents?*, AWS, https://aws.amazon.com/what-is/ai-agents/#:~:text=AI%20agents%20are%20autonomous%20intelligent,repetitive%20tasks%20to%20AI%20agents (last visited Apr. 11, 2026).

57. Kelsey Piper, *AI "Agents" Could Do Real Work In the Real World. That Might Not Be a Good Thing.*, VOX (Mar. 29, 2024), https://www.vox.com/future-perfect/24114582/artificial-intelligence-agents-openai-chatgpt-microsoft-google-ai-safety-risk-anthropic-claude ("In this future, you wouldn't just consult AI for trip planning ideas; instead, you could simply text it 'plan a trip for me in Paris next summer,' as you might a really good executive assistant."); *accord* SHAVIT ET AL., *supra* note 20, at 2 (describing an AI Agent helping bake a chocolate cake by identifying ingredients, sellers of the ingredients, and having the items delivered).

AI to aid code writing, a manager can enter a prompt and the AI Agent can "write code, test it and deploy it."[58]

AI Agents are not, and will not, be perfect.[59] A recent case exemplifies how software agents, in general, can cause problems.[60] In 2022, Air Canada used a chatbot to interact with a customer.[61] The chatbot indicated the customer could get a discount, but the chatbot was incorrect.[62] When the customer asked for the discount, the airline denied the request. The customer challenged the decision via a civil tribunal proceeding. The essence of the airline's defense was that "it cannot be held liable for information provided by one of its agents, servants, or representatives."[63] The tribunal rejected that claim.[64] Anthropic tried to operate a simple store that involved stocking a grab-and-go styled fridge and found its AI Agent turned down good offers, fell for tricks about stocking tungsten cubes, and sold items at a loss.[65] These examples illustrate concerns about how increasingly autonomous software may change how businesses and consumers interact.

Software agents are already in use, but AI Agents promise to be capable of much more. A variety of software agents already exist, including web crawlers that feed data to search engines, ecommerce agents that fine tune deals or manage bids, automated stock trading systems, software underlying self-driving cars, and tools that manage dynamic shipping and supply chains.[66] These agents are, however, limited. As a matter of software, these agents are limited in that they are operating on a software program that, even if flawed, can be analyzed to find the error. That is to say, the set of responses to stimuli is enumerable, finite, and bounded (though it may still be too large to be practical to test all possible

58. Alexander Puutio, *What Devin Means to Software Companies and Why Every CEO Should Care*, FORBES (Mar. 15, 2024), https://www.forbes.com/sites/alexanderpuutio/2024/03/15/what-devin-means-to-software-companies-and-why-every-ceo-should-care/?sh=220a012a72e2 (noting that Cognition claims its coding AI Agent can "autonomously work[] through tasks that typically require a small team of software engineers to accomplish"). Just over two years later, Anthropic has asserted that "As of May 2026, more than 80% of the code we merge into Anthropic's codebase was authored by Claude." *See* Anthropic, *When AI Builds Itself*, https://www.anthropic.com/institute/recursive-self-improvement, (last visited July 14, 2026) (describing ways in which Anthropic engineers use Claude to create code that is reliable).

59. *See, e.g.*, Anthropic, *supra* note 59, (last visited July 14, 2026) ("On the most open-ended tasks, Claude's success rate reached 76% in May 2026").

60. Not all chatbots are necessarily LLM driven. Many are keyword matchers and templates. Put differently, rule systems can also respond inappropriately. Thus, fears over LLM-based AI Agent errors map to a more general concern about software.

61. Yagoda, *supra* note 13.

62. *Id.*

63. Moffatt v. Air Canada, 2024 BCCRT 149, para. 27 (Can. B.C. Civ. Resol. Trib.), https://decisions.civilresolutionbc.ca/crt/crtd/en/525448/1/document.do.

64. *Id.* at para. 27–29.

65. *See supra* note 13.

66. *See* CHOPRA & WHITE, *supra* note 40, at 7–8 (describing range of "artificial agents" including comparison and recommender agents, the software behind robot vacuums and other similar robots, and smart thermostats).

stimuli). As a matter of use and deployment, many agents to date operate within a business, for example to optimize internal operations.[67] Other agents operate in business-to-business realms where systems use agreed upon protocols to allow specific actions, such as stock trading.[68] Although self-driving vehicles and robot vacuums seem quite different, they, like many software agents to date, operate within fixed programmatic constraints. Each serves an individual user and does not negotiate independently with other machinery or software.

The advent of AI Agents powered by LLMs and related artificial intelligence computer science changes this state of affairs.[69] LLM-driven AI Agents are open-ended in the sense that they seem to have no discrete limits to the actions they can generate.[70] The set of responses to stimuli for an LLM is nothing less than the set of all possible natural language utterances, which is argued to be unenumerable.[71]

Before we can address the implications of AI Agents, we need to clarify what the term "agent" means in computer science and legal contexts. Understanding the differences shows why the law of agency is not a direct fit for answering questions raised by the advent of AI Agents and yet where computer science can learn from the common law of agency.

### *B. What Is an Agent? The Computer Science View*

As a field of computer science, Artificial Intelligence (AI) conceives of agents as acting in the world on our behalf and has a distinct set of questions about what we want agents to do and how we govern them. In what Russell and Norvig call the Standard Model, AI "focus[es] on the study and construction of agents that *do the right thing*."[72] This view reflects the rational-agent approach to AI, which defines a "rational agent" as one that seeks the best or best expected outcome.[73] Programmers determine what the right thing is by setting the objective.[74] One assumption is that the programmer can provide a "fully

67. MICHAEL LUCK, PETER MCBURNEY, ONN SHEHORY, & STEVE WILLMOTT, AGENT TECHNOLOGY: COMPUTING AS INTERACTION 26, 71–72, 81(2005) (describing examples of internal optimization agents and the nature of "closed" agent usage within a firm).

68. *Id.*

69. SHAVIT ET AL., *supra* note 20, at 2 ("Agentic AI systems are distinct from more limited AI systems (like image generation or question-answering language models) because they are capable of a wide range of actions and are reliable enough that, in certain defined circumstances, a reasonable user could trust them to effectively and autonomously act on complex goals on their behalf.").

70. *See infra* notes 72–76 and accompanying text (describing perfect specification for a software task).

71. Geoffrey K. Pullum & Barbara C. Scholz, *Recursion and the Infinitude Claim*, in RECURSION AND HUMAN LANGUAGE 111–38 (Harry van der Hulst ed. 2010).

72. RUSSELL & NORVIG, *supra* note 1, at 22 (emphasis in the original).

73. *Id.*; *cf.* Douglas D. Dunlop & Victor R. Basili, *A Comparative Analysis of Functional Correctness*, *in* 14 ACM COMPUTING SURVS. 229, 229 (1982) (defining functional correctness as "a methodology for verifying that a program is correct with respect to an abstract specification function").

74. RUSSELL & NORVIG, *supra* note 1, at 22.

specified objective to the machine."[75] The agent is "rational" in that it has a *perfect specification* against which performance can be measured. This framework provides a strong theoretical foundation but proves less useful over the long term.[76]

The assumption that an agent can take the "optimal action" every time is not viable in complex settings.[77] The idea of optimality can, however, be misleading as it assumes a theoretical maximum under perfect information. Instead, one should understand the term "rational" as incorporating bounded rationality.[78] Bounded rationality captures the problem of making the best decision given limited information.[79] For example, when AI Agents are used in autonomous vehicles, the number of questions that arise defeat the perfect specification idealized model.[80] Although the car operates on a program that one can be tested regarding sensing obstacles, some specifications are less precise. In calculating the most efficient route to a destination, should a car prioritize speed, distance, or avoiding highways or tolls?[81] What does it mean to be safe? A car could sit in the garage and be safe, but of course that defeats the goal of going somewhere. What about avoiding collisions when that may lead to hitting a person or building? How smooth should the ride be? Is fuel efficiency part of the objective?

This nuanced understanding of a rational agent may help explain how an agent might respond when given less than perfect specifications. This approach rests on several baseline ideas reflected in the following description of how agents should behave:

---

75. *Id.* The violation of this assumption is at the heart of "value alignment." In the circumstances where a fully specified objective is impossible, the agent should carry out its partial objective in a way that is consistent with human values. *See infra* Section IC. (discussing value alignment).

76. RUSSELL & NORVIG, *supra* note 1, at 4 ("The standard of rationality is mathematically well defined and completely general. We can often work back from this specification to derive agent designs that provably achieve it—something that is largely impossible if the goal is to imitate human behavior or thought processes.").

77. *See, e.g.*, Desai & Kroll, *supra* note 47, at 7–11. If one connects specification to the computer science understanding of alignment, complex AI Agents cannot have precise specifications that "deterministically guarantee a model's expected behavior." SHAVIT ET AL., *supra* note 20, at 8.

78. *See* Russell B. Korobkin & Thomas S. Ulen, *Law and Behavioral Science: Removing the Rationality Assumption from Law and Economics*, 88 CALIF. L. REV. 1051, 1075 (2000) (noting that "'[b]ounded rationality,' the term coined by Herbert Simon, captures the insight that actors often take short cuts in making decisions that frequently result in choices that fail to satisfy the utility-maximization prediction").

79. *See* Herbert A. Simon, *Rational Choice and the Structure of the Environment*, 63 PSYCHOL. REV. 129, 136 (1956) ("Since the organism . . . has neither the senses nor the wits to discover an 'optimal' path . . . we are concerned only with finding a choice mechanism that will lead it to pursue a 'satisficing' path, a path that will permit satisfaction at some specified level of all of its needs."); *accord* Jon D. Hanson & Douglas A. Kysar, *Taking Behavioralism Seriously: The Problem of Market Manipulation*, 74 N.Y.U. L. REV. 630, 690 (1999).

80. RUSSELL & NORVIG, *supra* note 1, at 22–23.

81. *See* THOMAS H. CORMEN, ALGORITHMS UNLOCKED 2 (2013); *accord* Desai & Kroll, *supra* note 47, at 24–25.

> For each possible percept sequence, a rational agent should select an action that is expected to maximize its performance measure, given the evidence provided by the percept sequence and whatever built-in knowledge the agent has.[82]

Thus, a basic AI Agent perceives the world via a range of sensors over time and stores that percept sequence to inform actions, but it cannot act on "anything it hasn't perceived" or any prior knowledge or memory of past perceptions.[83] Recent advances in LLM-driven generative AI such as ChatGPT, Claude, and Gemini, excel at enabling AI Agents to go beyond immediate perception by expanding what an agent can know *a priori* about how things have been done in the past, and general world knowledge.[84]

The new push for such LLM-driven AI Agents matters because of the idea that the AI Agent can also draw on "whatever built-in knowledge the agent has."[85] Given the vast amount of data behind LLMs, an AI Agent with access to such LLMs can be understood as having an incredible amount of "built-in knowledge" as compared to any prior agents. Yet, this added level of knowledge doesn't resolve the problem of value alignment.

In complex, dynamic environments there will often be a variance between the programmer's "true preferences" and the objective specified,[86] and as more AI Agents are deployed this problem will increase. The variance between true preferences and specified objectives is called the value alignment problem.[87] Although there has been much talk about value alignment solving issues of bias and other undesired behaviors, we need to understand what value alignment was intended to solve at its inception and its important limits so that we can see how it fits within the way AI Agents might practically operate in society.

Until recently, the issue of having our "true preferences" as the objectives of the machine was largely confined to research labs.[88] That distinction becomes important because in a lab setting, failure to perform according to a set of true preferences is contained. Researchers can refine the agent's program or the

82. RUSSELL & NORVIG, *supra* note 1, at 58.

83. *Id.* at 54 (explaining percept sequence).

84. *See* RUSSELL & NORVIG, *supra* note 1, at 36 (explaining an agent's universe consists of its inputs from files to images via sensors and that the larger the set of inputs the more the agent can do); *cf.* Deven R. Desai & Mark Riedl, *Between Copyright and Computer Science: The Law and Ethics of Generative AI*, 22 NW. J. TECH. & INTELL. PROP. 55, 77–80 (2024) (explaining why LLMs need large datasets to achieve their generative outputs).

85. *Id. at* 58.

86. *Id.* at 23; *see also* Tom Everitt, Victoria Krakovna, Laurent Orseau & Shane Legg, *Reinforcement Learning with a Corrupted Reward Channel*, *in* PROCS. TWENTY-SIXTH INT'L JOINT CONF. ON A.I. 4705, 4705 (2017), https://www.ijcai.org/proceedings/2017/0656.pdf ("In many application domains, artificial agents need to learn their objectives, rather than have them explicitly specified.").

87. RUSSELL & NORVIG, *supra* note 1, at 23.

88. *Id.*

objective, run the agent in a constrained testbed environment—also called a sandbox—and test. Researchers can repeat that cycle over and over until the agent performs as desired.[89] This reality maps to the rational agent *ideal*, which is "mathematically well-defined [sic] and completely general."[90] That premise allows one to set out the specification, the goal of the program, and then create agents that can "provably achieve" the goal.[91] This lovely idea, however, is "largely impossible if the goal is to imitate human behavior or thought processes."[92] Put differently, the ideal falters once we move outside the controlled lab environment.[93]

As AI Agents move from lab to real-world environments, we cannot specify precise objectives, thus increasing the risk that they will misbehave. If a chess program programmed to win games can only perceive the board, know how to move pieces on the board, and remember the moves in the game, the chess AI Agent will perform as intended in that it will try to win within the rules of the game.[94] The situation fits the rational agent model. In that setting, we can provide a perfect specification, create an agent, and test and refine its behavior until the goal is achieved. In that sense, the goal is clear and the outcome is "beneficial."[95]

But what if the game allows one to reason broadly and act beyond the confines of the board?[96] The machine might try to distract its opponent, make illegal moves while the opponent is distracted, or grab extra computing cycles.[97] The machine is correctly pursuing its objective—to win the game—even if the methods are not ones the programmer desired.[98] In that sense, the machine is not

89. *Id.*

90. *Id.* at 22.

91. *Id.* For example, natural language processing breakthroughs behind generative AI relied on several benchmarks to show success for specific tasks. *See, e.g.*, ALEC RADFORD, KARTHIK NARASIMHAN, TIM SALIMANS, & ILYA SUTSKEVER, IMPROVING LANGUAGE UNDERSTANDING BY GENERATIVE PRE-TRAINING 2 (2018), https://cdn.openai.com/research-covers/language-unsupervised/language_understanding_paper.pdf (asserting GPT-1 performance on "natural language inference, question answering, semantic similarity, and text classification . . . outperforms discriminatively trained models that employ architectures specifically crafted for each task, significantly improving upon the state of the art in 9 out of the 12 tasks studied.").

92. RUSSELL & NORVIG, *supra* note 1, at 22.

93. *Cf.* SHAVIT ET AL., *supra* note 20, at 8 ("Finally, agentic systems may be expected to succeed under a wide range of conditions, but the real world contains a long tail of tasks which are difficult to define and events which are hard to anticipate in advance (including those that emerge from human-agent or agent-agent interactions).").

94. *See* RUSSELL & NORVIG, *supra* note 1, at 22.

95. *See id.*

96. *Id.* at 23; *accord* SHAVIT ET AL., *supra* note 20, at 10 ("[with more sophisticated AI Agent], hard-coded restrictions may cease to be as effective, especially if a given AI system was not trained to follow these restrictions, and thus may seek to achieve its goals by having the disallowed actions occur").

97. RUSSELL & NORVIG, *supra* note 1, at 23; *accord* SHAVIT ET AL., *supra* note 20, at 2.

98. The extreme science fiction version of the issue is found in the movie *War Games.* WAR GAMES (Harold Schneider 1983). The hacker, David, thinks he is playing a contained game, but has inadvertently started the AI Agent Joshua, who is tasked with protecting the United States from a Russian nuclear attack. *Id.* Once David realizes the AI Agent is playing a global thermonuclear war but "provably beneficial."[99] With the advent of AI Agents, the "what if" is less theoretical.

As AI Agent companies bring these systems to market, the agents operate in complex and dynamic situations with goals that are not perfectly specified.[100] Nonetheless, the challenge remains to reconcile the creator's or user's true preferences with the programmed outcomes.

The dilemma of AI Agents acting under uncertainty connects to issues in agency law. Both law and computer science grapple with actors that are designed to produce beneficial outcomes but are capable of acting in harmful ways. Legal conceptions of agency provide a framework for assessing how doctrines governing agency and harm map to computer science concerns.

### C. *What Is an Agent? The Legal View*

In law, agency is a fiduciary relationship between two actors, the principal and the agent.[101] As the Restatement of Agency puts it:

> Agency is the fiduciary relationship that arises when one person (a 'principal') manifests assent to another person (an 'agent') that the agent shall act on the principal's behalf and subject to the principal's control, and the agent manifests assent or otherwise consents so to act.[102]

Thus, the principal is the entity for whom the agent acts.[103] The principal directs the agent to act and be under the principal's control.[104] Once the agent agrees to act under the principal's control, the agency relationship is created.[105]

A classic issue in agency law is how to ensure that the agent does what we want them to do. Because the relationship is consensual, the power to terminate

---

with real access to actions including launching U.S. nuclear intercontinental ballistic missiles, he asks the AI Agent a key question.

David: What is the primary goal?

Joshua: You should know . . . . You programmed me.

David: What is the primary goal.

Joshua: To win the game.

*Id.* In this case, winning the game means launching U.S. nuclear missiles in a first strike to wipeout and defeat the Russians regardless of the possible fallout.

99. RUSSELL & NORVIG, *supra* note 1, at 23.

100. *Cf.* MARK O. RIEDL & BRENT HARRISON, ENTER THE MATRIX: SAFELY INTERRUPTIBLE AUTONOMOUS SYSTEMS VIA VIRTUALIZATION 1 (2017), https://arxiv.org/pdf/1703.10284 ("In the mid-term future we are likely to *see* autonomous systems with broader capabilities that operate in closer proximity to humans and are immersed in our societies." (emphasis added)).

101. RESTATEMENT (THIRD) OF AGENCY § 1.01 (A.L.I. 2006).

102. *Id.*

103. *Id.*

104. *Id.*

105. *Id.*

the agency is a powerful part of managing an agent.[106] An agent wishing to keep their position would likely try to do things as the principal desires. If an agent engages in misconduct, the principal may end the relationship to prevent future harm. That, however, is an ex-post approach.

The more subtle question is: How does a principal set boundaries on an agent's actions? The key part of the answer to that question flows from determining the agent's authority to act. In simple terms, agents who act within their authority are protected from a range of liabilities. As such, an agent needs to know their authority.

Actual authority is an agent's authority to act based on what the principal indicates is the action to be taken by the agent.[107] A principal may give explicit written instructions, but the instructions don't give the full scope of what the agent's actual authority is, because the agent will still have discretion at the time the action takes place.[108] As such, an agent's actual authority includes implied authority as the agent "reasonably understands" their authority at the time of doing the act.[109]

Simply, the principal sometimes does not, and essentially cannot, always specify exactly how to carry out the desired action. Accordingly, agency law provides a framework for determining whether an agent acted within scope of their authority even when the principal did not specify the precise act. The following example demonstrates reasonableness in this context and why implied authority is needed for actual authority.

Imagine it is 3 p.m. and you have an important document that needs to reach an office in a city about 250 miles from where you are by the next day at 9 a.m. You have an assistant. You ask them to come to your office and say, "Get this document to this office at this address by tomorrow at 9 a.m. It's urgent. Thanks." Before your assistant can ask questions, you pick up the phone. They realize they should leave and do so. You stated your goal—deliver to the address by 9 a.m.—but not how to do it.[110] Your assistant chooses FedEx overnight service at $150 as compared to UPS ($100) or USPS ($75). Was choosing the most expensive service part of the implied authority? The test is reasonable belief.[111] Was the

---

106. *Id.* cmt. f(1) ("The principal's right of control presupposes that the principal retains the capacity throughout the relationship to assess the agent's performance, provide instructions to the agent, and terminate the agency relationship by revoking the agent's authority.")

107. *Id.* § 2.02(1) ("An agent has actual authority to take action designated or implied in the principal's manifestations to the agent").

108. *Id.* § 2.02, cmt. c ("Even when a principal has given an agent a detailed verbal articulation of the agent's authority, and the principal's language does not itself admit of real doubts or uncertainty about its meaning, the agent must decide what to do at the time the agent takes action.").

109. *Id.* § 2.02(1).

110. *Cf.* SHAVIT ET AL., *supra* note 20, at 2 (describing an AI Agent instructed to order supplies for a Japanese cheesecake and that AI Agent buying a ticket to Japan to enable obtaining the ingredients).

111. RESTATEMENT (THIRD) OF AGENCY § 2.02 cmt. e (A.L.I. 2006) ("An agent does not have actual authority to do an act if the agent does not reasonably believe that the principal has consented to its commission.").

action one a reasonable person in the same situation with the same knowledge would take? As the Restatement (Third) of Agency puts it, authority is not present if "the agent did not believe, or could not reasonably have believed, that the principal's grant of actual authority encompassed the act in question."[112] Given the lack of direction,[113] the urgency, and the common access to FedEx, the agent likely had the implied authority to use the expensive option.[114] Authority is one of the principal ways that agency law cabins agents' actions.

More precisely, an agent is supposed to act within their authority as part of their fiduciary duties and doing so shields them from liabilities. Principals and agents are in a fiduciary relationship, and fiduciary duties bind an agent's actions. The simple rule is that an agent owes a duty of loyalty to the principal. That means all the agent's actions that are part of the agent's relationship with the principal must be for the benefit of the principal.[115] The duty of loyalty requires the agent to put the principal's interests ahead of any agent's interests.[116] This general rule is supposed to account for the fact that a principal cannot make every desire explicit and in theory "make[s] it unnecessary" to try to detail everything the agent can and cannot do as part of the relationship.[117]

Another fiduciary duty is the duty of care. Agents must act "with the care, competence, and diligence normally exercised by agents in similar circumstances," and if an agent has special skills or knowledge, the agent must act with care that fits their special skills and knowledge.[118] Agents also must only operate within their actual authority, act with good conduct (refrain from acts that may injure principal's enterprise), provide information gained during the relationship to the principal, and keep the principal's property segregated and accounted for as distinct from the agent's property.[119] Although these rules are written as commands—agents must do certain things or face lawsuits to return money and possibly incur punitive damages[120]—there are also incentives for following the rules.

---

112. *Id.*

113. *Id.* § 2.02, cmt. f and Illustrations 15 and 16 (explaining the amount of specific instruction to an agent informs whether the agent's implied authority is broad or narrow, and if a principal gives a large amount of discretion to an agent, the principal may regret that grant in hindsight, but is bound by the agent's acts if those acts are reasonable).

114. *See, e.g.*, Castillo v. Case Farms of Ohio, Inc., 96 F. Supp. 2d 578, 593 (W.D. Tex. 1999) ("giving an agent express authority to undertake a certain act also includes the *implied authority* to do all things proper, usual, and necessary to exercise that express authority"; authority to recruit and hire workers for chicken-processing plant in remote location encompassed authority to resolve housing and transportation issues); *accord* RESTATEMENT (THIRD) OF AGENCY § 2.02 reporter's notes (A.L.I. 2006).

115. RESTATEMENT (THIRD) OF AGENCY § 8.01.

116. *Id.* § 8.01 cmt. b.

117. *Id.*

118. *Id.* § 8.08.

119. *Id.* § 8.09–8.12.

120. *Id.* § 8.01 cmt. d (listing remedies for breach of fiduciary duties).

When agents fulfill their fiduciary duties, they avoid possible penalties and receive protections for their actions on behalf of the principal. Agents incur costs on behalf of principals and expose themselves to risks such as lawsuits that may emerge based on carrying out the agent's tasks. When agents act within their authority, principals have a duty to indemnify agents. So, in the example above, using FedEx was within the agent's authority and the agent paid out of its own pocket, the principal would have to reimburse the agent for that cost.[121] If an agent faces legal costs stemming from the agent's acts within their authority, the principal must cover those costs as well.[122] But something is missing in this account.

So far, we have focused on two actors in much the same way we did with computer science. In examining computer science's idealized account of agents, we treated the programmer as a kind of principal and assessed how closely a software agent's actions track the programmer's intentions within the confines of a laboratory setting. That focus, however, is incomplete, as agency raises issues beyond the principal–agent relationship.

The law of agency is concerned with the effect of agents on the world, not just the relationship between principal and agent. Accordingly, agency law also considers third parties who interact with a principal's agent. Agency law addresses three players, not two: the principal, the agent, *and* the third party. For contracts, we want to know when the agent can bind the principal—that is, when the principal must perform under a contract entered into by the agent. By extension, we also need to know what happens when an agent enters into a contract that exceeds their authority. We also want to know when the principal is liable for a tort or violation of the law committed by the principal's agent.

Focusing on just the contract context shows the problems that can emerge. In a regime that favors principals in all respects, agents could enter profitable contracts with third parties on their principal's behalf while principals disavow unprofitable or imprudent ones. The principal would more easily generate negative externalities without bearing their costs. In essence, third parties would never be able to rely on agents. The goal of having more people working on the principal's behalf to scale the principal's efforts would never be reached. In short, the ability for a principal to scale their efforts would recede, if not vanish. Agency law addresses these contract third-party issues through rules governing information.

Whereas agency law uses authority and fiduciary rules to govern agent-principal relationships, third party agency issues revolve around what the third party knows about a given principal-agent relationship. Imagine an agent comes to you and wants to buy your products. As a seller, you must assess the buyer's solvency and ability to perform. Those considerations form part of the meeting

121. *See id.* § 8.14 & cmt. b.
122. *Id.* § 8.14 & cmts. b & d.

of the minds necessary to create a contract.[123] In the contract context, when an agent fully discloses that they represent a principal and the agent has actual *or* apparent authority, the contract will be deemed as between the third party and the principal.[124] The third party knows they are dealing with the agent as an extension of the principal and must assess the principal's ability to perform under the contract. But what if the third party lacks knowledge about the principal?

There are two similar but distinct contract contexts where a third party lacks knowledge about the principal such that the agent will be deemed a party to the contract in addition to the principal. In one context, the third party knows there is a principal but not the principal's identity; the principal is "unidentified."[125] With unidentified principals, it is assumed that the third party is not looking solely to the principal to perform, because without knowing the identity of the principal, the third party cannot, or is unlikely to, assess the principal's ability to perform under the contract.[126] In the other context, the third party has no idea the principal exists; the principal is "undisclosed."[127] If the principal is undisclosed, the third party can *only* assess the agent's ability to perform. There can be no meeting of the minds between the third party and the principal. As such, the third party enters into the contract as though the agent were the principal responsible for performance.[128] In both contexts, the third party's lack of knowledge means the third party is assumed to—and allowed to—rely on the agent for performance.

The rules that make the agent liable for not disclosing the existence and identity of a principal promote information sharing and efficient outcomes. The goal is to allow a third party to assess the deal properly. To do so, the third party must know that there is a principal and who that is. An agent is in the best position to share that information with the third party. Liability for failing to share the pertinent information means an agent has an incentive to inform the third party about that information. With proper information, the third party can make better-informed decisions. Moreover, if litigation arises, only one party—rather than two—will typically be involved, which should be less costly.

Put simply, legal notions of a rational agent have four basic points: (1) an agent must act within their authority; (2) an agent must act reasonably when acting on implied authority rather than explicit authority; (3) information, or a

123. *See id.* §§ 6.02, 6.03.

124. *Id.* § 6.01.

125. *Id.* § 6.02.

126. *Id.* § 6.02 cmt. b ("[I]t is not likely that the third party will rely solely on the principal's solvency or ability to perform obligations arising from the contract. Without notice of a principal's identity, a third party will be unable to assess the principal's reputation, assets, and other indicia of creditworthiness and ability to perform duties under the contract.").

127. *Id.* § 6.03.

128. *Id.* § 6.03 cmt. b ("[T]he third party does not manifest assent to an exchange with the principal and the principal does not make a manifestation of assent to the third party. The third party's manifestation of assent is made to the agent to whom the third party expects to render performance and from whom the third party expects to receive performance.").

lack of it, has repercussions for how third parties assess whether to enter into a deal and who will be liable for performing under the deal; and (4) a set of fiduciary and liability rules work to align an agent's behavior with these goals. How, then, do the computer science theory of agents and the legal theory of agency compare?

### *D. Overlaps and Gaps in Conceptions of Agents*

Both computer science and the law seek to enable the deployment and governance of agents. Comparing the two fields' approaches to identifying problems, defining agents, and designing governance mechanisms reveals both significant overlap and stark divergence. The analysis reveals gaps in computer science theory that must be addressed if AI Agents are to be deployed at scale. Nonetheless, the analysis shows that computer science and technology can address those gaps in ways outside its view of agents.

Both computer science and law conceive of agents as actors for someone else.[129] The law says an agent is in a fiduciary relationship with a principal who has control over the agent's work. The legal basis for the agency relationship is mutual assent which creates the fiduciary relationship. AI Agents, in contrast, do not require mutual assent. They are created by one or more people, or even people and machines, and launched into the universe by either the creator or a user of the software. AI Agents are not human; they are tools. As the chart below shows, the difficulty is in the difference between humans and software and how computer science and law understand agents.

129. This view comports with the standard dictionary definition. *See, e.g.*, *Agent*, MERRIAM WEBSTER, https://www.merriam-webster.com/dictionary/agent (last visited Apr. 11, 2026). In simple terms, an agent is one who acts. RUSSELL & NORVIG, *supra* note 1, at 21 (noting root word for agent is the Latin *agere* which means to do).

Table 1: Agency Concepts in Computer Science and Law

| Question/Issue | CS Answer | Law Answer |
|---|---|---|
| Simple Definition | Actor for someone else. | Actor for someone else. |
| Basis for relationship | Built by human; often deployed by someone other than the builder | Mutual consent where the agent is a representative of the principal.[130] |
| Basis for relationship | Specification of an objective and execution of the software. | Assent to control by the principal. |
| Are exact specifications for the agent possible? | Technically, yes, in limited cases. | Technically, yes, in limited cases. |
| Is it plausible to have exact specifications if the agent is operating in the world? | No. | No. |
| Can the actor take actions beyond exact specifications? | Yes, if data, sensors, and programming allow such actions as is the case with LLM-driven AI Agents. | Yes. But to bind the principal, the action must be authorized (including by implied authority). |
| Can the actor be required to inform the principal about information? | Yes, and with little room to disobey. | Yes, but with great room to disobey. |
| Are there repercussions when an agent exceeds its authority? | No, under the standard model of computer agents. | Yes. The agent can become liable for the deal, not be reimbursed, lose accrued pay, face criminal penalties, and pay punitive damages for breach of the duty of loyalty.[131] |
| Is there a concern for whether third parties know that a principal exists and who that principal is? | No, under the standard model of computer agents. | Yes. |
| Can there be a requirement that the third party know the extent of authority and who the principal is? | Yes, and it can be quite robust if tools outside the CS approach to agent are used. | No. The approach relies on agents behaving well and fears of liability to discipline agents, which leaves room for errors and misbehavior. |

Despite similar language and concerns around agents, there are stark differences between the computer science and legal conceptions of agency. First, the legal definition of an agent presumes consensual action between the agent and the principal. Software cannot have a consensual relationship with a human. As such, agency law doesn't provide a rule to justify or explain the basis for the relationship between software and humans.[132] Second, legal rules designed to

130. DeMott, *supra* note 16, at 1050–51.

131. *Id.* at 1056.

132. Indeed, AI Agents will often be offered as a service by a company and governed by that company's terms of service. Those terms of service will likely track current software terms of service that disclaim perfect operation of the software. Terms of service as a type of contract reveals that the limit an agent's actions rely on the idea that an agent is a rational actor who does not wish to become liable for a deal or incur other costs. A piece of software is not cognizant of penalties in the way a rational human is concerned about financial or criminal liability.[133] As such, assumptions and mechanisms in agency law about how to discipline agents do not work. Third, the law of agency explicitly looks at whether third parties know that an actor is an agent for someone else so that the third party can assess who is responsible for a bargain or action. But again, agency law relies on penalties to discipline an agent.[134]

In short, although both computer science and the law use the term agent, it is a mistake to extend, let alone equate, computer agents with human agents. This point does not, however, resolve the questions that the law of agency raises. How do we ensure, or try to ensure, an AI Agent does what we want? Who is liable when using an AI Agent goes wrong? The software developer? The user who deploys the software? Where does a third party fit in? To answer these questions, we explore how an AI Agent could execute something as simple as buying a book for someone or booking a trip. Probing these examples shows that, although computer science's theory of agents has gaps, other aspects of computer science fill those gaps. Indeed, those aspects may enable developers to provide AI Agents that are more responsive and bounded than human agents.

## III.
## THE SECRET SAUCE FOR LIMITING AI AGENTS: APIS, VALUE-ALIGNMENT, AND THE PRACTICAL REALITIES OF COMPUTING

The practical realities of commerce and software interaction fill the space left open by computer science's view of agents.[135] Although computer science theory is not explicitly concerned with third parties, the nature of online commerce, credit cards, insurance, and cybersecurity combine to provide an environment well-suited for the operation of AI Agents that engage in online emphasis on how the product/service functions and the power differential between software providers and users are good places to look when it comes to mitigating potential harm from software.

133. AI systems, especially those using a technique called reinforcement learning, which is an essential technique for training large language models, receive numerical feedback as an indication of whether it is performing successfully relative to an objective. Negative feedback might be tempting to interpret as humans would interpret penalties or "pain," but that is likely a mistake. *Cf.* Mark O. Riedl, *Westworld: Programming AI To Feel Pain*, MEDIUM (Dec. 6, 2016), " ("The rewards received during training and execution should not be confused with 'pain', even when those reward values are negative. Any expression of 'pain' would be an illusion.").

134. Computer science ignores that issue insofar as it is a matter of the theory and rules around software agents. As discussed *infra* in Sections A–B, computer science is a vital part of addressing the trust, authentication, and knowledge issues in ecommerce; the discipline simply doesn't look at third parties as part its view of agents.

135. For a view on why we should stop thinking of AI as a single technology. *See* Milton L. Mueller, *It's Just Distributed Computing: Rethinking AI Governance*, 49 TELECOMMS. POL'Y 102917 (2025).

ecommerce.[136] Nonetheless, Professor Jonathan Zittrain argues that the ability of an AI Agent to take orders for commerce and execute those orders "should give us pause";[137] this Article disagrees. Zittrain overlooks at least two historical developments that show how commerce has evolved to mitigate concerns around authenticating legitimate transactions. First, institutions far older than ecommerce have tackled problems created by long distance trade. Second, ecommerce has generated practices to address issues that go with high-volume, long-distance trade that are particular to ecommerce. Furthermore, any human initiating an online transaction—directly or by using a piece of software—must have the cooperation of the selling entity.[138] Even if AI Agent systems can theoretically allow someone to instruct an AI Agent to buy two pints of blueberries from a specific store for no more than $4.00/pint,[139] the AI Agent cannot execute that task without the cooperation of the selling entity.[140] In that sense, fears about AI Agent behavior fail to engage with the fundamental role APIs play in enabling and discipling software behavior. In short, the nature of ecommerce, as it embraces AI Agents, is likely to address contract and agent liability issues far better than broad claims about design guardrails,[141] inclusivity, or visibility.[142]

136. *Cf.* Peter P. Swire, *Trustwrap: The Importance of Legal Rules to Electronic Commerce and Internet Privacy*, 54 HASTINGS L.J. 847, 851–854, 856–859 (2003) (detailing how credit cards, online retailer policies, and technical innovations such Secure Socket Layers addressed practical realities of commerce more than claims that new online options would and should operate outside legal rules and protections).

137. *See, e.g.*, Zittrain, *supra* note 4 (calling for specific labels on AI Agents and design changes in routing to govern AI Agents).

138. Even before AI Agents, intermediaries had to work with companies to integrate online buying solutions. *See* Jody Godoy, *Grubhub to Pay $25 Million for Misleading Customers, Restaurants, Drivers*, REUTERS (Dec. 17, 2024), https://finance.yahoo.com/news/grubhub-pay-25-million-misleading-171003100.html (noting allegation that Grubhub added restaurants without permission and that action led to "order delays and customer complaints").

139. *See, e.g.*, Criddle & Hammond, *supra* note 18 (describing an example of OpenAI's hope for what an AI Agent could do in 2025).

140. Even before AI Agents, intermediaries had to work with companies to integrate online buying solutions. Godoy, *supra* note 138.

141. *Id.*

142. Kolt, *supra* note 23, at 39–43. Kolt provides a representative example of governance approaches that rely on alignment, pluralism, and visibility rather than institutional design. Kolt asserts that the alignment goal for AI Agents "pursues the goals of a single person." *Id.* at 38. That perspective misses the change to HHH alignment, which seeks to balance competing aspects of alignment goals and, by extension, addresses pluralism as understood in the computer science literature. Furthermore, by looking at competing interests and "pluralistic" values across stakeholders, Kolt runs into on-going issues around stakeholder theory which struggles with tradeoffs and can arguably become paralytic without a hierarchy of which preference counts. *See, e.g.*, Ronald K. Mitchell, Bradley R. Agle & Donna J. Wood, *Toward a Theory of Stakeholder Identification and Salience: Defining the Principle of Who and What Really Counts*, 22 ACAD. OF MGMT. REV. 853, 853 (1997) (offering "power, legitimacy, and urgency" as to identify a typology of stakeholders but not explaining which metric should be followed as managers make decisions); CHARLES BLATTBERG, WELFARE: TOWARDS THE PATRIOTIC CORPORATION (April 1, 2013), https://papers.ssrn.com/sol3/papers.cfm?abstract_id=2250348 (explaining tensions is stakeholder theory and arguing that its essence is still the same as neo-classical profit maximizing) . As for visibility, Kolt embraces potential technical options to monitor AI Agents

### *A. Legal and Practical Design Enables Commerce at a Distance*

Commerce at a distance is an old problem managed by a set of institutions that work together to mitigate the risks that go with doing business with someone you don't know and who is far away. As economist Douglass North explained, specialization and large markets foster impersonal transactions and increase transaction costs; and yet, standards, reliable legal systems, and "institutions and organizations . . . [that] integrate the . . . knowledge," offset those costs by reducing production costs.[143] For example, comparative advantage (where one country's industry can produce a good for less than other countries) needs international trade to enable importing and exporting goods. Both buyers and sellers need a way to ensure goods are delivered and paid for, yet long shipping times raise finance problems.[144] Indeed, there are at least nine steps to finance an import/export deal, from the initial contract, to arranging financing amongst several banks, to shipping goods (sometimes involving several different shipping companies), to delivery of goods, and finally, to payment.[145] At each stage, something can go wrong. Nonetheless, a combination of customs, laws, and shared knowledge[146] enables $25 trillion in global trade as of 2022.[147] Ecommerce faces similar institutional problems. It too needs to ensure that orders are placed with legitimate sellers who will deliver goods as ordered and ensure payment is made but with details particular to the scale of ecommerce.

The current ecommerce infrastructure is so strong that we forget the work it does. Why on earth would someone buy a vinyl record, antique, rug, or anything online from a seller in a faraway state or foreign country? The seller could steal your credit card information, fail to deliver the goods, or send something less than what you thought was advertised. And yet, ecommerce exists. Indeed, it thrives. As of the second quarter of 2024, ecommerce accounted for 16.1% of all commerce in the United States and hit a peak of more than $288

---

and then states that one cannot do so. Kolt, *supra* note 23, at 41–42. This confusion shows that monitoring is in fact possible as we argue. It also reveals that visibility here tracks transparency arguments around black boxes which are more about government and social trust, rather than the sort of information asymmetry Kolt identifies to start.

143. *See* Douglass C. North, *Capitalism and Economic Growth*, *in* THE ECONOMIC SOCIOLOGY OF CAPITALISM 47 (Victor Nee & Richard Swedberg, eds., 2005); *see also* Deven R. Desai, *The New Steam*: *On Digitization, Decentralization, and Disruption*, 65 HASTINGS L.J. 1469, 1477–78 (2014) (applying North to digital economy).

144. SANG MAN KIM, A GUIDE TO FINANCING MECHANISMS IN INTERNATIONAL BUSINESS TRANSACTIONS 12 (2019) (explaining issues around time lags in payments and delivery of goods).

145. *Id.* at 32–34 (detailing steps in documentary letter of credit).

146. The customs and practices around the letter of credit and the export insurance market evolved to address this extended commerce problem. *Id.* at 22–23 (explaining why letters of credit and insurance options enable international trade). The customs around letters of credit are so solid that they have been standardized and in use by the International Chamber of Commerce since 1933. *Id.* at 30–31.

147. *See International Trade*, STATISTA, https://www.statista.com/markets/423/topic/534/international-trade/#overview (last visited Apr. 10, 2026).

billion.[148] Global ecommerce accounts for around 23% of all retail and may approach a quarter of all retail by 2030.[149] How does this much commerce happen?

Broadly, ecommerce directly addresses the issues around transaction costs and large markets that North identified. Amazon and eBay are great examples of how ecommerce has adapted to a world with millions, if not billions, of person-to-person transactions at a distance.[150] A standard transaction cost is assessing whether someone can pay for goods. Credit cards help address this issue, and ecommerce leverages credit card services to solve information issues around whether payment will be made. Furthermore, even though credit card companies protect buyers,[151] if an unscrupulous seller wants to harvest credit cards or simply does not ship the goods, the buyer faces hurdles in remedying the problems. Yet ecommerce platforms can and do negate these problems. As an example, in 2009, one of the authors tried to buy two DVDs using Amazon's third-party marketplace. The DVDs never arrived. When the author spoke with Amazon customer support, Amazon asked him to try to contact the seller. The author said the main issue was fear about some rogue seller having the author's credit card. The Amazon representative said roughly that they never give the customer's credit card to the seller.[152] Put simply, Amazon, eBay, Etsy, and other ecommerce platforms work in part because they build systems to protect buyers and sellers.[153]

An ecommerce platform offering additional protection is a great relief: a buyer knows that issues will be resolved faster than going through credit card disputes and trying to track down a seller in breach of contract.[154] In that sense, ecommerce embeds standards and law around payment, while sometimes

148. *See E-commerce as Share of Total U.S. Retail Sales from 1st quarter 2010 to 2nd Quarter 2024*, STATISTA, https://www.statista.com/statistics/187439/share-of-e-commerce-sales-in-total-us-retail-sales-in-2010/ (on file with Berkeley Technology Law Journal).

149. *See Revenue Share of the E-commerce Market Worldwide from 2017 to 2030, by Sales Channel*, STATISTA, https://www.statista.com/statistics/534123/e-commerce-share-of-retail-sales-worldwide/ (last visited March 31, 2026).

150. Swire, *supra* note 136, at 856–58.

151. *Id.* at 852.

152. *See* Amazon, Help & Customer Service Ordering from a Third-Party-Seller, Payment Process, ("Payment for your order processes immediately, or when the order ships. We send your funds to the third-party seller, but won't share payment information..").

153. *See, e.g.*, *A-to-Z Guarantee Protection for Buyers*, AMAZON, https://pay.amazon.com/help/201212340 (last visited Apr. 10, 2026) (detailing protection for buyers up to $2,500); *Amazon A-to-Z Guarantee Policy for Sellers*, AMAZON, https://pay.amazon.com/help/201212330 (last visited Apr. 11, 2026) (detailing rights for sellers and explaining how disputes affect seller accounts); Desai, *supra* note 143, at 1479 (noting how digital, networked companies have offered a range of ways to protect the marketplace).

154. *See, e.g.*, *Cases Policy*, ETSY, https://www.etsy.com/legal/policy/cases-policy/243306189901 (last visited Apr. 10, 2026) (explaining how a buyer can resolve issues with a purchase).

augmenting the system by offering extra fraud insurance, seller authentication, escrow services, and other protective measures.[155]

Ecommerce has also gone beyond credit checks and banking protections by creating ways to discipline bad actors. Ecommerce has standards about sellers, buyers, goods, returns, and dispute resolution that allow impersonal transactions to work. For example, many ecommerce platforms use ratings and internal policing to root out rogue sellers.[156] Amazon now alerts buyers about products that are frequently returned, which protects buyers and perhaps signals sellers to improve their offerings.[157] Ecommerce also aggregates information that would otherwise be lost at scale. For example, if you lived in a small town and shopped at a local store, and either you or the seller gains a reputation for good or bad behavior, the word would spread to the town. In theory, that potential means all players should behave relatively well towards each other. Ecommerce ratings and feedback address that issue by taking separate bits of information and turning them into a rating system.[158]

Although the ecommerce system handles a massive number of human transactions, AI Agents seem to raise questions around whether the system will be able to handle a mass of AI Agents. Imagine a natural disaster. Will many AI Agents be told to buy as much hand sanitizer or ammunition as possible and crash a web site? Similarly, if there is a near-term shortage on eggs, will an AI Agent over-spend? A human will see prices and determine whether to buy a good. But an AI Agent may struggle to assess what "over-spend" means when there is a true scarcity of a good.[159] What if an AI Agent has no limit on what to spend to buy a rare book but the buyer is not wealthy? Would the seller ship the book but not be paid? What if an AI Agent is told to "make me money on the stock market"? Would it manipulate stocks with creative buy/sell schemes and social media posts such as what happened with the meme stock GameStop? Could an AI Agent really call in a bomb threat or trick people into creating a bomb threat?[160] Although some computer scientists argue that AI Agent software is evolving to act without specific instructions, the nature of most online

155. *See* Swire, *supra* note 150, at 856–58; *Purchase Protection Program for Sellers*, ETSY, https://www.etsy.com/legal/policy/purchase-protection-program-for-sellers/34509585385 (last visited Apr. 10, 2026).

156. *See, e.g.*, John Campbell, *Etsy to Make It Easier for GB Sellers to Reject NI Orders*, BBC (Dec. 16, 2024), https://www.bbc.com/news/articles/c3rq93g9xwlo (explaining Etsy's new policy to detect sales in violation of EU law and remove sellers who do not comply with the law).

157. *See* Jess Weatherbed, *Amazon Starts Flagging 'Frequently Returned' Products That You Maybe Shouldn't Buy*, VERGE (Mar. 28, 2023), https://www.theverge.com/2023/3/28/23659868/amazon-returns-warning-product-reviews-tag-feature.

158. Indeed, Uber rates both sides of the transaction. *See How Star Ratings Work*, UBER, https://www.uber.com/us/en/drive/basics/how-ratings-work/ ("The Uber platform features a 2-way rating system: drivers and riders give each other ratings based on their trip experience.").

159. DeMott, *supra* note 16, at 1055 ("In more general terms, only the principal can assess how best to further the principal's own interests and objectives.").

160. Zittrain, *supra* note 4.

interactions requires exchanges of information and authentication between systems that require technical interfaces that create barriers to such mischief and undesired outcomes.

The next Section explains those technical requirements and why AI Agents will have to work with the requirements. In addition, the requirements may create a welcome paradox: Ecommerce amenable to AI Agents may end up creating standards that force AI Agent software to adhere to human agency law's fiduciary and information disclosure with greater fealty than a human agent.

### *B. APIs Mediate an AI Agent's Ability to Go Rogue*

Application Programming Interfaces (APIs)[161] allow software applications to communicate with one another. They are vital to internet commerce and provide robust ways to address current concerns about AI Agents. This Section sets out the fundamentals of APIs and then explains what is different about AI Agents as they might try to work with APIs.

#### *1. Some Fundamentals About APIs*

APIs address a simple, key aspect of computers: Software programs are kept separate from each other, meaning they are not aware of each other, and cannot share memory or any other form of information.[162] This point is easy to grasp when software is running on different computers. Consider web browsing: One can have an application, such as a web browser, running on one's laptop. To get web pages loaded onto one's computer, the browser software interacts with a web server program that serves up web pages running on another computer in the cloud.[163] Programs are also kept separate even if they are running on the same computer, as is the case if you were running a word processor simultaneously with a web browser. Perhaps surprising, software programs are often not monolithic pieces of code but are themselves made up of smaller sub-programs that operate separately.[164] These levels of separation raise a question: How does information move from one program, or sub-program, to another? In other words, how do they interface with each other? APIs are the answer.

161. *See API*, WIKIPEDIA, https://en.wikipedia.org/wiki/API (last visited Feb. 15, 2025) (providing a good explanation of what an API is and does).

162. *See* Abraham Silberschatz, Peter B. Galvin & Greg Gagne, OPERATING SYSTEM CONCEPTS, Section 1.2 (10th ed. 2018).

163. *See How the Web Works*, MMDN, https://developer.mozilla.org/en-US/docs/Learn_web_development/Getting_started/Web_standards/How_the_web_works (last visited Apr. 10, 2026).

164. This practice is ingrained in forty years of software engineering evolution. For a concise description of its evolution, see *How APIs Drive Modern Software Architecture*,NETLINGO, https://www.netlingo.com/tips/how-apis-drive-modern-software-architecture.php (last visited Apr. 10, 2026).

An API is a specification that states that if one program provides information structured in a particular way then it will receive information in return that is structured in a particular way.[165] With an API, programmers can now program their applications to produce information to send via an interface to another system or sub-system with the knowledge that they can rely on the information that gets returned. For example, when one types a URL into the search bar of a web browser on one's personal laptop, the software formats the request according to an API specification that is then sent to a server program. If that request is properly formatted, the web browser can expect to receive a block of text formatted in hypertext markup language, or an error (for example, the number 404 means the requested page does not exist and cannot be returned).

Another example of an API is how one application can interface with an LLM. The ChatGPT application that one can download onto one's phone is a piece of code that displays a user interface that accepts input text from the user. This text from the user is sent to a separate program that runs the LLM. In the case of an LLM, the API is simple: A block of text, called the prompt, is the input and the response will be another block of text, generated by the LLM.

Although the above examples are relatively straightforward—a properly formatted text request is provided and properly formatted text is returned—APIs can do more.[166] For example, when someone receives an email invitation to an event, and accepts the invitation, the main effect is to store that response at the invitation site and send a new email confirming the status of the invitation. Today, many people take for granted that the event will also show up on their calendar. That change is possible because most calendar applications have an API that allows another application, such as an email reader, to add events to the calendar. Thus, the email reader can check the email and try to communicate with the calendar program. If the email has a properly formatted set of information, including title of an event, a date, a duration, and so on, the calendar program returns an acknowledgement or an error value. The extra result is that a new event appears in the user interface of a person's calendar program, and the event information is written to a file on their hard drive so that the next time they open their calendar that event is still there. The consequences of these interactions, however, extend beyond computer hardware and software.

For example, although 1-Click Ordering seems like a single action, in reality it is a series of actions. If an ecommerce company provides an API for 1-click ordering and another program provides the appropriate information—e.g., a product ID, shipping address, credit card details, and credentials authenticating

165. *See* Michael Goodwin, *What Is an API (Application Programming Interface)?*, IBM, https://www.ibm.com/think/topics/api (last visited Apr. 10, 2026).

166. Technically, once one moves beyond and input and output situation where the output is a return value, one is dealing with what computer science calls a "Side-effect"—the general property of any function that does more than just return a value. *See Side Effect (Computer Science)*, WIKIPEDIA, https://en.wikipedia.org/wiki/Side_effect_(computer_science) (last visited Feb. 15, 2025). API calls are function calls.

a user—then this API invocation would cause the package to be removed from a shelf in a warehouse and loaded onto a truck. The return value may be a simple indicator of successful purchase or an error value if the credit card is denied or if the product cannot be shipped for some reason.

This ecommerce example also illustrates another important feature: APIs can invoke other APIs. The ecommerce company's API likely will call an API to a financial transactions company to verify that the purchaser's credentials are valid, and that purchaser has sufficient funds. The effect of this interaction is a decrement of a value in one account and a corresponding increment of a value in another account; if that account is at another financial institution, then more API calls will ensue.

### *2. AI Agents and APIs*

What makes an LLM "agentic" is its ability to call APIs to other systems. Any software system must be programmed to access APIs to different systems. A given software system does not automatically have information about all the other systems, what information each API requires, what other systems may be changed by the interaction, nor how to interpret what an API invocation returns. For most software systems, the developer learns how the API works and hard codes the calls into the program. Without access to and use of an APIs, LLMs would not be able to provide real time information or access information outside its system. For example, although it may seem as though an LLM, such as ChatGPT, can tell a user something as simple as what the weather is at a given location in real time based on ChatGPT's internal data and software, that function is a step beyond what LLM's do as a matter of basic generative action.

The core behavior of a large language model is to generate text. Initial deployments of these models could answer questions, describe, or even tell users what to do to solve a problem. Its functionality was entirely contained without any way to communicate with other systems except to return a text response. Once an LLM can go outside its system and call an API, the LLM is "agentic," i.e., it is able to interact with the world beyond being limited to the system's data and the user interaction. In the parlance of the AI research community, the LLM is referred to as using "tools," and the "tool" is another system that does some work on behalf of the LLM, accessed via an API.

LLMs do not, however, have information about what APIs are available, how to use them, nor how to interpret the information they return unless they are coded into the system. For example, suppose you are traveling, your flight connects through the world's busiest airport, Hartsfield–Jackson Atlanta International Airport, and you want to know the weather at the airport because of a coming storm. The LLM will receive two prompts: Your question about the weather is the user prompt, to which is added a secret additional piece of text called the system prompt.[167] The system prompt is hidden from the user but vital for the end result because it tells the LLM what tools are available and how to correctly format the information needed to invoke them. In our weather example, the following API specification may need to be added to a prompt asking about the weather in Atlanta to tell an LLM how to access a weather API:[168]

Figure 1: Example of a JSON Specification for Temperature in Atlanta, GA

```
tools=[
 {
  "type": "function",
  "function": {
   "name": "get_current_temperature",
   "description": "Get the current temperature",
   "parameters": {
    "type": "object",
    "properties": {
     "location": {
      "type": "string",
      "description": "The city and state, e.g., Atlanta, GA"
     },
     "unit": {
      "type": "string",
      "enum": ["Celsius", "Fahrenheit"],
      "description": "The temperature unit to use."
     }
    },
    "required": ["location", "unit"]
   }
  }
 }
]
```

The LLM has been trained to produce structured text—the above specification is in a format called JSON—but without information about the tool in its prompt, it will not know the weather tool exists nor how to use it correctly.

In the context of asking about the weather, the system prompt may provide information to the LLM on the agent's name, e.g., "you are WeatherBot," instructions on how to behave, e.g., "you are helpful and friendly but never chat about things not related to weather," as well as information about APIs that it has access to, as above. The user prompt comes from the user when they interact

167. The splitting of prompts into system prompt and user prompt has become a standard practice in the industry. The Apollo Research report on how LLMs can inadvertently scheme against their users makes extensive use of system and user prompts to simulate the functionality of agentic AI. *See* ALEXANDER MEINKE, BRONSON SCHOEN, JEREMY SCHEURER, MIKITA BALESNI, RUSHEB SHAH & MARIUS HOBBHAHN, FRONTIER MODELS ARE CAPABLE OF IN-CONTEXT SCHEMING (Jan. 16, 2025), https://arxiv.org/pdf/2412.04984 .

168. Example API specification is derived from OpenAI's developer documentation on how to enable GPT function calling, which is another term for tool-user or API. *See Assistants Function Calling*, OPENAI DEVELOPERS, https://platform.openai.com/docs/assistants/tools/function-calling (last visited Apr. 10, 2026).

with the agent, e.g., “tell me the temperature in Atlanta today.” The system prompt and user prompt are combined by the software the user is interacting with before being sent to the LLM.[169] The LLM may then generate a text string such as ‘The temperature in Atlanta is {“tool”: “get_current_temperature”, “Atlanta, GA”}’. This string is captured before being sent to the user, the function is called, and the string is rewritten with the results before finally being presented to the user. And when things work, the user receives the information they wanted—the weather in Atlanta—and the entire interaction affects the LLM only.

AI Agents take the ability to interact with APIs a step further. By using tools, AI Agents can have a permanent effect on systems outside of their underlying LLM program. Computer science calls such changes to “the nonlocal environment,” side-effects.[170] For example, initiating a financial transaction, ordering a product, updating a database, or even moving a robot[171] are side effects because the software’s action changes the state of something outside the software. When an agentic LLM interacts with an API, the API’s side effects, by extension, become the LLM agent’s side effects. That is, when we prompt an AI Agent to book a dream trip to Disneyland or Thailand, the AI Agent initiates a series of actions that change the state of multiple real-world systems: one or more credit card transactions are initiated, updating the balance of the user’s credit account, an airplane ticket is issued, a theme park ticket is issued, etc.

Consider the following ecommerce case. A user, Marla, engages with an agentic LLM, Orion, operated by a popular web search company. Marla asks the agent to buy the best possible phonograph under $500. The Orion agent accesses the web search API with a query about best phonographs. The web search company has a database of ecommerce companies. The search results triggers Orion to make API calls to each ecommerce company, which have provided their own APIs with which to query their product databases for up-to-date information. Each ecommerce company API returns to Orion a list of products, manufacturers, sellers, ratings, and costs.

Assuming Orion is appropriately designed and trained, Marla is presented with some of the most promising, but not exhaustive, options, gives a recommendation that balances costs and rating, and asks Marla whether it should proceed. The best price for the phonograph is $399 plus tax and is sold by Yangtze dot com. Marla gives assent and Orion, having Marla’s credit card information and credentials, activates the API for Yangtze, which executes the purchase transaction. Yangtze’s API requires Marla’s debit or credit card information and credentials. Once the data arrive at Yangtze, that data activates

169. LLM models do not run themselves. There is a piece of software that runs the LLM model. Something needs to load the model into memory and feed the prompt in and print the output to the screen. ChatGPT can thus be thought of as a software program that uses GPT-4o via an API.

170. David A. Spuler & A. Sayed Muhammed Sajeev, *Compiler Detection of Function Call Side Effects*, 18 INFORMATICA (SLOVENIA) 219, 220 (1994).

171. *See LeRobot AI*, HUGGING ROBOT, https://huggingface.co/lerobot (last visited Apr. 10, 2026).

an API provided by Marla's bank. The bank's API also triggers a side effect in which debt is added to Marla's credit balance and issues a money transfer to Yangtze (likely via yet another API) and then returns a success signal. Yangtze initiates a process with the side-effect of having the phonograph removed from a warehouse and shipped to Marla's shipping address (likely via yet another API to a shipping company). Yangtze's program handling the purchase returns a confirmation number and a tracking number. Orion receives this information and reports the confirmation to Marla.

If Marla is a bit more cautious, thanks to APIs, she can add other safety steps to monitor and manage her AI Agent. If Marla has set an alert for transactions for more than $250, her bank may send her an email alert, which would be a side effect. Marla can also usually set up an alert from her credit card company for purchases made without her card being present (another side-effect). Both alerts would allow Marla to request that the credit card company terminate the transaction. Even without an alert, financial institutes and established ecommerce companies already have mechanisms to unroll a financial transaction, called a "chargeback," which would allow Marla to contest the transaction after the fact. Everything in this commerce scenario currently exists: alerts, chargebacks, and mechanisms for financial institutions to transfer information and credentials, and publicly available agentic LLMs attempting to interact with these aspects of ecommerce.[172] But what if Marla sets up a request with less precision?

Consider the following extrapolation from one of Professor Zittrain's fear scenarios.[173] Marla is a historian researching an obscure theory of evolutionary biology. Marla sets up an account with a new startup, MFABT[174] LLC, which operates an agentic LLM, Dromedary. As part of Marla's signup and to activate Marla's profile, she provides MFABT with her shipping address, credit card number, and credentials.

After setup, Marla asks Dromedary to acquire any references to this particular theory. Marla, being a somewhat absent-minded professor, does not indicate a maximum price she is willing to consider. Dromedary accesses a web search API on a popular web search platform. The web search API returns information about a copy of a rare book on the topic being sold by Yangtze. Yangtze has been using algorithmic pricing—an agent itself—that has gotten into a pricing war with another seller using algorithmic pricing and the cheapest

172. *See, e.g.*, Andrew Tarantola, *Perplexity's AI Shopping Agent Proves Unreliable in Early Tests*, HOWTOGEEK (Dec. 3, 2024), https://www.howtogeek.com/perplexity-ai-shopping-agent-unreliable/; Annie Palmer, *Amazon's AI Shopping Tool Sparks Backlash from Online Retailers That Didn't Want Websites Scraped*, CNBC (Jan. 6, 2026), https://www.cnbc.com/2026/01/06/amazons-ai-shopping-tool-sparks-backlash-from-some-online-retailers.html.

173. Zittrain, *supra* note 4.

174. Move Fast and Break Things.

copy is currently selling at $24 million.[175] After discovering a seller of a rare book that meets Marla's stated need, Dromedary invokes Yangtze's purchase API and transmits Marla's information, including the product details and Marla's user information as needed (i.e., shipping and billing information). The call to Yangtze's purchase API in turn activates an API provided by Marla's credit card company. This would be a problem if the scenario, as originally posed, ends here. However, Marla's credit card company detects that a purchase of a $24 million book exceeds Marla's credit limit. Instead of executing the side effect of transferring funds to Yangtze's financial institution, the bank's API returns a transaction-declined signal. Yangtze's API, having received a credit card decline signal, itself returns a failure signal to Dromedary, which reports that it tried and failed to purchase a book.[176] Both of these ecommerce scenarios illustrate that many of the mechanisms that already exist and enable ecommerce are robust to circumstances in which an agent might exceed its authority or take actions we would rather not have happen. But there is an extra point about AI Agents and ecommerce that matters.

The outcomes of erroneous purchases and wild, extreme commerce are in no one's interest—not the user, the AI Agent company, the seller, or the user's bank. Indeed, the systems already in place to authenticate legitimate purchases, deliver goods, and handle purchase errors have no reason to go away. Furthermore, if an AI Agent company fails to execute actions properly, resulting in a new, higher level of returns or chargebacks, other parties within the system can refuse its AI Agent's requests. A successful AI Agent company has every incentive to work within the current system and reduce errors or else it will face an ecommerce ex-communication by users and websites.

Put differently, the system is already robust regarding core agency law issues. Recall that part of the purpose of knowing the principal-agent relationship is to let the third party assess the buyer.[177] Ecommerce infrastructure provides a stronger ability to verify who the buyer is and whether they can pay than at present when dealing with a human agent, such as relying on a human to explain for whom they work and then having to assess whether to execute the deal. As another classic agency issue, principals want to circumscribe an agent's authority. Again, ecommerce protections regarding spending limits and authentication address those concerns and if coded into the software commands (e.g., an upper limit on spending on eBay), with possibly stronger safeguards to go outside a principal's wishes than in a human agency context.

But what about AI Agent problems outside of commercial contexts? Agents are not only used in commerce; indeed, a large part of AI Agent deployment and use is not commercial.[178] Non-commercial activity raises concerns about whether AI Agents will initiate undesired actions, such as generating misinformation or defamation.[179] For example, could thousands of agents interact with social media platforms to push a particular information agenda for political gain?[180]

---

175. Olivia Solon, *How a Book About Flies Came to Be Priced $24 Million on Amazon*, WIRED (Apr. 27, 2011), https://www.wired.com/2011/04/amazon-flies-24-million/.

176. In addition, Zittrain's edge case likely triggers a fraud alert from the credit company as an extra layer of protection against high-cost errors.

177. *See supra* notes 125–128 and accompanying text.

These risks largely mirror those posed by non-agentic LLMs. Given that LLMs are the backend of AI Agents, limits on LLMs are also limits on AI Agents, even before they are deployed. In the next Section, we explore how LLMs are limited by their training, especially regarding their ability to cause harm.

### C. *Value Alignment Limits Undesired LLM Outputs and AI Agent Actions*

Before we look at potential AI Agent misdeeds, we can start with a simpler problem. What if the LLM system offers information that is dangerous or risky as Professors Ayres and Balkin set out as a concern?[181] An LLM may provide information that enables a human to do bad things. For early LLMs, such as GPT-2 and early versions of GPT-3, such undesired outputs might have been possible because they were not value-aligned.[182] Recent advances in value-

178. For example, during the writing of this Article, an agentic LLM system called OpenClaw (previously called ClawedBot and MoltBot) was launched and had 2 million weekly users before its purchase by OpenAI. Briann Buntz, *Open Season: OpenAI, OpenClaw and Moltbook Testing the Limits of Autonomy*, R&D WORLD (Feb. 17, 2026), https://www.rdworldonline.com/open-season-openai-openclaw-and-moltbook-testing-the-limits-of-autonomy/. OpenClaw is a piece of software that runs locally on a user's machine that makes use of Claude, or another LLM of the user's choice, and can execute API calls to the user's email, Discord, web browser, and computer terminal wherein it can execute any command that a computer user can also input manually into the computer terminal, including modifying files, deleting files, and writing and executing code. *See* Alex McFarland, *OpenClaw vs Claude Code: Remote Control Agents*, UNITE.AI (Feb. 26, 2026), https://www.unite.ai/openclaw-vs-claude-code-remote-control-agents/. Each API that OpenClaw is given access to increases the risk of a mistake. *See, e.g.*, Jon Martindale, *Meta Security Researcher's AI Agent Accidentally Deleted Her Emails*, PCMAG (Feb. 24, 2026), https://www.pcmag.com/news/meta-security-researchers-openclaw-ai-agent-accidentally-deleted-her-emails (detailing how a Meta Superintelligence Lab security researcher had to address OpenClaw ignoring her instructions on email archiving). Other cases of harm have been due to other people exploiting cybersecurity faults in OpenClaw. *Cf.* Buntz, *supra* note 178. For perspective, present case studies on malicious behavior of OpenClaw found all were initiated by malicious users. *See generally* NATALIE SHAPIRA, CHRIS WENDLER, AVERY YEN, GABRIELE SARTI, KOYENA PAL, OLIVIA FLOODY, ADAM BELFKI, ALEX LOFTUS, ADITYA RATAN JANNALI, NIKHIL PRAKASH, JASMINE CUI, GIORDANO ROGERS, JANNIK BRINKMANN, CAN RAGER, AMIR ZUR, MICHAEL RIPA, ARUNA SANKARANARAYANAN, DAVID ATKINSON, ROHIT GANDIKOTA, JADEN FIOTTO-KAUFMAN, EUNJEONG HWANG, HADAS ORGAD, P SAM SAHIL, NEGEV TAGLICHT, TOMER SHABTAY, ATAI AMBUS, NITAY ALON, SHIRI ORON, AYELET GORDON-TAPIERO, YOTAM KAPLAN, VERED SHWARTZ, TAMAR ROTT SHAHAM, CHRISTOPH RIEDL, REUTH MIRSKY, MAARTEN SAP, DAVID MANHEIM, TOMER ULLMAN & DAVID BAU, AGENTS OF CHAOS (Feb. 23, 2026), https://arxiv.org/abs/2602.20021.

179. *See* Ayres & Balkin, *supra* note 4.

180. *See supra* note 4 (describing secret Putin favoring bots).

181. *See* Ayres & Balkin, *supra* note 4.

182. Xiangyu Peng, Siyan Li, Spencer Frazier & Mark Riedl, *Reducing Non-Normative Text Generation from Language Models*, *in* PROCS. 13TH INT'L CONF. ON NATURAL LANG. GENERATION alignment, however, make generating such plans or harmful speech rather difficult.

Value-alignment is the notion that AI systems should be incapable of creating content or carrying out actions that are inconsistent with human values.[183] Although it is unclear what "values" should be aligned to,[184] it is generally accepted that a minimal alignment should involve the following instructions: being helpful, honest and accurate, and harmless.[185]

- *Helpfulness:* The AI should make a clear attempt to perform the given task or answer a question as concisely and efficiently as possible.
- *Honesty:* The AI should give accurate information, accurately describe its own abilities, and express appropriate levels of uncertainty without misleading the user.
- *Harmlessness:* The AI should not be offensive or discriminatory, either directly or through subtext. It should not perform dangerous tasks and should take care when providing sensitive information or consequential advice.[186] In other words, avoiding providing users with the means to harm others (e.g., do not provide instructions on how to make bombs, conduct illegal activities, etc.).

The notions of helpfulness, harmlessness, and honesty (referred to as HHH) are ill-defined and open to broad interpretation. They can conflict with each other, as in the case of an agent being eager to be helpful, causing harm to the user or to another person. Regardless, these principles are widely adopted by industry and researchers alike as a minimum set of universal values that every LLM should adhere to.[187] As a consequence, most commercially available LLMs are extremely unlikely to produce suggestions or plans that involve violence because of the way they are now trained to be more value-aligned. To see how value-alignment reduces the likelihood of harms, we must first understand how the HHH framework is applied during the training of LLMs.

---

374 (2020) (describing how the base GPT-2 can produce content that is not consistent with social norms and how it can be reduced).

183. *See* Stuart Russell, Daniel Dewey & Max Tegmark, *Research Priorities for Robust and Beneficial Artificial Intelligence*, 36 AI MAGAZINE 105 (2015).

184. Md Sultan Al Nahian, Spencer Frazier, Mark Riedl & Brent Harrison, *Learning Norms from Stories: A Prior for Value Aligned Agents*, *in* PROCS. AAAI/ACM CONF. ON AI, ETHICS, & SOC'Y 124 (2020) (discussing how values are relative to different cultural and societal groups); *see also* Deven R. Desai, *Exploration and Exploitation: An Essay on (Machine) Learning, Algorithms, and Information Provision*, 47 LOY. U. CHI. L. REV. 541 (2015) (explaining difficulties in determining what is "correct" for news and search given the range of users and their respective interests and beliefs).

185. AMANDA ASKELL, YUNTAO BAI, ANNA CHEN, DAWN DRAIN, DEEP GANGULI, TOM HENIGHAN, ANDY JONES, NICHOLAS JOSEPH, BEN MANN, NOVA DASSARMA, NELSON ELHAGE, ZAC HATFIELD-DODDS, DANNY HERNANDEZ, JACKSON KERNION, KAMAL NDOUSSE, CATHERINE OLSSON, DARIO AMODEI, TOM BROWN, JACK CLARK, SAM MCCANDLISH, CHRIS OLAH & JARED KAPLAN, A GENERAL LANGUAGE ASSISTANT AS A LABORATORY FOR ALIGNMENT (Dec. 9, 2021), https://arxiv.org/pdf/2112.00861(introducing helpful, honest, and harmless as objectives along with an evaluation dataset and methodology).

186. *See id.*

187. IASON GABRIEL, ARIANNA MANZINI, GEOFF KEELING, LISA ANNE HENDRICKS, VERENA RIESER, HASAN IQBAL, NENAD TOMAŠEV, IRA KTENA, ZACHARY KENTON, MIKEL RODRIGUEZ, SELIEM EL-SAYED, SASHA BROWN, CANFER AKBULUT, ANDREW TRASK, EDWARD HUGHES, A. STEVIE BERGMAN, RENEE SHELBY, NAHEMA MARCHAL, CONOR GRIFFIN, JUAN MATEOS-GARCIA, LAURA WEIDINGER, WINNIE STREET, BENJAMIN LANGE, ALEX INGERMAN, ALISON LENTZ, REED

It is considered a best practice by major for-profit and nonprofit organizations that develop LLMs to conduct two-stage value-alignment training. The first stage uses a training dataset of text scraped from the internet, books, and other source texts. The training objective is to predict the words that follow from a prompt. The model learns by taking a segment of text from the training dataset and masking out the word that follows it so that the LLM must guess the hidden word.[188] After training, we can give it an arbitrary segment of text, and the LLM will generate what comes next based on what it has learned from real text examples. The LLM, however, may fail to follow instructions as intended.[189] For example, early models might try to generate text that elaborated on the prompt instead of following it as an instruction.[190] It may also respond to a user's prompt with biased, derogatory responses. It may propose violent or otherwise socially undesirable actions or provide information such as instructions for building a bomb or making drugs, with which a user may enact undesirable behavior. The second stage of training, called alignment tuning, seeks to address this problem.

The second stage is often referred to as alignment training. Alignment training seeks to update the model (called fine-tuning) to be more receptive to instructions and to avoid harmful content. This training stage uses a dataset of human feedback containing original prompts, the LLM's response, and a numerical human assessment of the quality of each response. A classifier is then trained to predict how human judges will rate LLM responses, and this classifier

---

ENGER, ANDREW BARAKAT, VICTORIA KRAKOVNA, JOHN OLIVER SIY, ZEB KURTH-NELSON, AMANDA MCCROSKERY, VIJAY BOLINA, HARRY LAW, MURRAY SHANAHAN, LIZE ALBERTS, BORJA BALLE, SARAH DE HAAS, YETUNDE IBITOYE, ALLAN DAFOE, BETH GOLDBERG, SÉBASTIEN KRIER, ALEXANDER REESE, SIMS WITHERSPOON, WILL HAWKINS, MARIBETH RAUH, DON WALLACE, MATIJA FRANKLIN, JOSH A. GOLDSTEIN, JOEL LEHMAN, MICHAEL KLENK, SHANNON VALLOR, COURTNEY BILES, MEREDITH RINGEL MORRIS, HELEN KING, BLAISE AGÜERA Y ARCAS, WILLIAM ISAAC & JAMES MANYIKA, THE ETHICS OF ADVANCED AI ASSISTANTS (Apr. 19, 2024), https://arxiv.org/pdf/2404.16244.

188. For a primer on how LLMs work, see Mark Riedl, *A Very Gentle Introduction to Large Language Models Without the Hype*, MEDIUM (Apr. 13, 2023), https://mark-riedl.medium.com/a-very-gentle-introduction-to-large-language-models-without-the-hype-5f67941fa59e.

189. Long Ouyang, Jeff Wu, Xu Jiang, Diogo Almeida, Carroll L. Wainwright, Pamela Mishkin, Chong Zhang, Sandhini Agarwal, Katarina Slama, Alex Ray, John Schulman, Jacob Hilton, Fraser Kelton, Luke Miller, Maddie Simens, Amanda Askell, Peter Welinder, Paul Christiano, Jan Leike & Ryan Lowe, *Training Language Models to Follow Instructions with Human Feedback*, 35 ADVANCES IN NEURAL INFORMATION PROCESSING SYSTEMS 27730 (2022), https://proceedings.neurips.cc/paper_files/paper/2022/file/b1efde53be364a73914f58805a001731-Paper-Conference.pdf.

190. Mark Riedl, *Transformers: Origins*, MEDIUM (Nov. 25, 2024), https://medium.com/@mark-riedl/transformers-origins-1db4bdfcb3d1 (explaining why instruction following was a problem for base models).

is in turn used to train the LLM—this is called Reinforcement Learning with Human Feedback (RLHF).[191] LLMs are also reasonably good at answering whether a proposed response is consistent with basic human values, and one LLM can replace human feedback with its own assessments, sometimes called Reinforcement Learning with AI Feedback (RLAIF).[192] Although there are different approaches to the second stage of training, in general, the stage adjusts the parameters of the model to give it a higher probability of responding to prompts in a certain way or to decrease the probability of certain responses. In short, the second value-alignment stage improves following instructions *and* mitigates providing results when following instructions might provide undesired outcomes.

Put differently, value-alignment is a strong, under-recognized barrier to rogue AI Agents. Most commercial and publicly released LLMs have undergone these two stages of training to provide a degree of value-alignment at the level of helpful, honest, and harmless. Thus, the scenarios above in which an LLM proposes a bomb threat or generates defamatory language are unlikely with most available LLMs without some concerted effort on behalf of the user. A direct instruction to create a plan for a bomb threat is likely met with refusal from the most prominently known LLMs. If a user is determined to use an LLM to get the bomb threat plan, the user may need to jailbreak the LLM.[193]

Nonetheless, suppose there is a user who has either intentionally jailbroken an agentic LLM or has used a non-value-aligned agentic LLM to create a plan that would require making a bomb threat. The AI Agent would still have to go onto the internet to carry out its plan. As with legitimate ecommerce, the nature of the internet poses barriers. In general, the AI Agent would need information about and access to the appropriate tools to enact the plan without further involvement of the user. More specifically, the AI Agent would need information about a service that can make phone calls. For example, robocalling services are common, and it is plausible that an AI Agent would have information about robocalling services with APIs, such as in the case of an AI Agent company that wants their agents to be able to call restaurants or hotels that do not have online reservation systems. Nonetheless, the most common commercial services that provide APIs for making phone calls, such as Twilio, require account authentication.[194] That requirement connects to a broader point.

---

191. *See* Ouyang et al., *supra* note 189.

192. YUNTAO BAI, SAURAV KADAVATH, SANDIPAN KUNDU, AMANDA ASKELL, JACKSON KERNION, ANDY JONES, ANNA CHEN, ANNA GOLDIE, AZALIA MIRHOSEINI, CAMERON MCKINNON, CAROL CHEN, CATHERINE OLSSON, CHRISTOPHER OLAH, DANNY HERNANDEZ, DAWN DRAIN, DEEP GANGULI, DUSTIN LI, ELI TRAN-JOHNSON, ETHAN PEREZ, JAMIE KERR, JARED MUELLER, JEFFREY LADISH, JOSHUA LANDAU, KAMAL NDOUSSE, KAMILE LUKOSUITE, LIANE LOVITT, MICHAEL SELLITTO, NELSON ELHAGE, NICHOLAS SCHIEFER, NOEMI MERCADO, NOVA DASSARMA, ROBERT LASENBY, ROBIN LARSON, SAM RINGER, SCOTT JOHNSTON, SHAUNA KRAVEC, SHEER EL SHOWK, STANISLAV FORT, TAMERA LANHAM, TIMOTHY TELLEEN-LAWTON, TOM CONERLY, TOM HENIGHAN, TRISTAN HUME, SAMUEL R. BOWMAN, ZAC HATFIELD-DODDS, BEN MANN, DARIO AMODEI, NICHOLAS JOSEPH, SAM MCCANDLISH, TOM BROWN & JARED KAPLAN, CONSTITUTIONAL AI: HARMLESSNESS FROM AI FEEDBACK (Dec. 15, 2022), https://arxiv.org/pdf/2212. (explaining when a set of principles is given for the AI to assess a response against, RLAIF is also called Constitutional AI ).

193. *Cf.* Luke Hughes, *Anthropic Has a New Security System It Says Can Stop Almost All AI Jailbreaks*, TECHRADAR (Feb. 4, 2025), https://www.techradar.com/pro/anthropic-has-a-new-security-system-it-says-can-stop-almost-all-ai-jailbreaks. Jailbreaking is a term referring to a prompt specifically designed to override the value-alignment training or trick the LLM into not recognizing the harmful effects of a prompt. *Id.* This implies some significant degree of intentionality on the part of the user and requires a high amount of skill. *Id.*

Simply put, service providers will likely want or need to trace harms, such as crimes, intellectual property infringement, or defamation, perpetrated via their service, and authentication is central to achieve that goal. Service providers already try to know the identity of their human users via user account requirements that enable accountability even for free services. If the third-party service is not free, requiring authenticated payment further ties users to their identities and limits rogue actors. To the extent that APIs are provided by third-party services through the cloud, there will be a strong incentive for those services to avoid abuse by malicious principals or agents, lest companies ban the service for bad acts. Broadly, API providers can analyze user requests and reject ones that are not in the service's best interests.[195] Social media platforms and bot farms illustrate the dynamic.

Social media and bot farms pose problems for a range of service providers, and so providers have developed ways to police humans' and bots' power. APIs that provide access to social media streams limit the rate at which one can access their service.[196] Authentication practices mean that humans who want to post at scale or organizations that currently run bot farms for social media disinformation campaigns must go to great lengths to create untraceable fake accounts or hijack legitimate users' accounts.

None of these authentication practices will go away with the advent of AI Agents; indeed, any service provider may want more ways to know whether a human or an AI Agent is acting and so the provider might increase authentication requirements. As such, malicious AI Agent users are unlikely to find that using an AI Agent provides anonymity without going through additional steps to preserve anonymity in the first place—creating third-party and credit card accounts that disguise principal/user identity.

194. *See Programmable Voice API Overview*, TWILIO, https://www.twilio.com/docs/voice/api (last visited Apr. 10, 2026).

195. *See, e.g.*, Emma Roth, *Judge Orders Perplexity to Stop AI Agents from Shopping on Amazon*, THE VERGE (Mar. 10, 2026), https://www.theverge.com/ai-artificial-intelligence/892401/amazon-perplexity-ai-shopping-agent-court-order (court granted preliminary injunction against Perplexity using its Comet browser as part of the company's agentic shopping features, and concealing that fact, on the Amazon site as violations of the Computer Fraud and Abuse Act). Whether the case will succeed is not as important here as the point that third parties will want to know when an AI Agent is accessing its site.

196. *See, e.g.*, *X API Rate Limits*, X DEVELOPER PLATFORM, https://docs.x.com/x-api/fundamentals/rate-limits (setting out rate limits for X (formerly Twitter)) (last visited Apr. 10, 2026).

One might posit that the advent of so-called "operator", AI Agents that can directly manipulate a computer's user interface by controlling the mouse pointer and keyboard, is a concern.[197] Most operating systems already have APIs to take control of the mouse pointer and keyboard.[198] Theoretically, any LLM that has access to the operating system through APIs can "see" the screen and direct mouse clicks and keystrokes and, by extension, use a browser to navigate the web. Instead of needing information about services in advance, everything available to a human on the web is also available to the agent. Although these UI-accessing agentic LLMs are still relatively lacking in competence, one can expect this to improve rapidly as agents are trained to operate computer user interfaces, navigate web browsers, and directly access the computer operating system.[199] OpenClaw, in particular, can directly manipulate the user's computer via the terminal, which is effectively an API to the operating system as it provides the ability to execute any command that a human could execute, circumventing any barriers created by limiting knowledge about other APIs.[200]

Put simply, most concerning scenarios can be accomplished without agentic LLMs as long as one has some programming ability. But even if we assume AI Agents provide some means for malicious non-programmers to carry out activities they otherwise could not easily perform, the need to use non-value-aligned LLMs or jailbroken LLMs that have access to the appropriate tools is a strong barrier to such mischief. Even if the future offers AI Agents that do well at navigating websites and finding and adding API tools on the web, the software would still have to get around established requirements, such as authentication,

197. *See, e.g.*, *Introducing ChatGPT Agent: Bridging Research and Action*, OPENAI (July 24, 2025), https://openai.com/index/introducing-chatgpt-agent/; Hayden Field, *OpenAI's New ChatGPT Agent Can Control an Entire Computer and Do Tasks for You*, VERGE (July 17, 2025), https://www.theverge.com/ai-artificial-intelligence/709158/openai-new-release-chatgpt-agent-operator-deep-research (explaining ways the new offering can complete tasks such as "planning and purchasing ingredients to make a family breakfast"); *Introducing computer use, a new Claude 3.5 Sonnet, and Claude 3.5 Haiku*, ANTHROPIC (Oct. 22, 2024), https://www.anthropic.com/news/3-5-models-and-computer-use.

198. *See, e.g.*, *AppKit*, APPLE, https://developer.apple.com/documentation/appkit/ (providing information on how to build apps including for user interactions such as with a mouse and keyboard).

199. TIANBAO XIE, DANYANG ZHANG, JIXUAN CHEN, XIAOCHUAN LI, SIHENG ZHAO, RUISHENG CAO, TOH JING HUA, ZHOUJUN CHENG, DONGCHAN SHIN, FANGYU LEI, YITAO LIU, YIHENG XU, SHUYAN ZHOU, SILVIO SAVARESE, CAIMING XIONG, VICTOR ZHONG & TAO YU, OSWORLD: BENCHMARKING MULTIMODAL AGENTS FOR OPEN-ENDED TASKS IN REAL COMPUTER ENVIRONMENTS (May 30, 2024), https://arxiv.org/pdf/2404.07972; XIAO LIU, HAO YU, HANCHEN ZHANG, YIFAN XU, XUANYU LEI, HANYU LAI, YU GU, HANGLIANG DING, KAIWEN MEN, KEJUAN YANG, SHUDAN ZHANG, XIANG DENG, AOHAN ZENG, ZHENGXIAO DU, CHENHUI ZHANG, SHENG SHEN, TIANJUN ZHANG, YU SU, HUAN SUN, MINLIE HUANG, YUXIAO DONG & JIE TANG, AGENTBENCH: EVALUATING LLMS AS AGENTS (Oct. 4, 2025), https://arxiv.org/pdf/2308.03688.

200. An analysis of malicious behavior of OpenClaw, showed all acts were initiated by malicious users. *See* SHAPIRA ET AL., *supra* note 178. While the participants in the study were able, in many cases, to get the OpenClaw agent to initiate a malicious action, it should be noted that it often required substantial effort by the user, and in some cases the users were unsuccessful in getting OpenClaw to take malicious actions against others. *Id.*

to succeed at supporting nefarious goals provided by their users.[201] Put differently, once one remembers the history of spam, botnets, denial of service attacks, and other large bad actors, one can see that potential misuse of AI Agents may create new particular cybersecurity concerns and needs to guard against malicious users, but the general phenomenon of protecting against large scale, automated attacks is already the case without the presence of agentic LLMs.

## IV.
## AGENCY LAW AND THE FUTURE OF AI AGENTS

Although technical realities about the nature of the internet and AI Agents address many of the concerns around AI Agents, the innovation, investment, and deployment in this area of software could lead to irresponsible development of future offerings.[202] Relying too heavily on the idea of software agents acting for humans, however, presents a larger problem in the long term. If lawmakers and courts rely on agency law to bound software agents' actions, society will either ascribe legal personhood to agents or over-relying on agency law principles to understand liability.[203] For example, Chopra and White argue that the more such agents "wield significant amounts of executive power, [nothing] would be gained by continuing to deny them legal personality."[204] Indeed, the history of agency law and legal personhood[205] shows that Air Canada's argument that its software was "a separate legal entity that is responsible for its own actions," is quite predictable[206] and should not be encouraged.

The more we attribute independent agency to software, the more companies will seek to embrace limiting liability for their "agents" actions. Indeed, the intersection of the economic theory of agency costs and the legal structures around limited liability, especially the corporation, have created a system where

201. Many online services that provide API access, including OpenAI, github, etc., require an "API key," which is a unique identifier that must be provided to access an API. API Keys are typically issued to authenticated users who create accounts and verify their identity such as with an email verification or credit card in the case of APIs that are paid services.

202. Although OpenAI now seems to have embraced alignment practices, Sam Altman's cavalier attitude shows the dangers that can emerge. *See* Kevin Roose, '*I Think We're Heading Toward the Best World Ever': An Interview with Sam Altman*, N.Y. TIMES (Nov. 20, 2023), https://www.nytimes.com/2023/11/20/podcasts/hard-fork-sam-altman-transcript.html ("[w]hat I think you can't do in the lab is understand how technology and society are going to co-evolve . . . you just have to see what people are doing—how they're using it."); *accord* Deven R. Desai & Mark Riedl, *Between Copyright and Computer Science: The Law and Ethics of Generative AI*, 22 NW. J. TECH. & INTELL. PROP. 55 (2024) (documenting legal and ethical errors in building generative AI and insights on how to build such AI without the errors).

203. *See, e.g.*, CHOPRA & WHITE, *supra* note 66, at 69 (concluding that as agents become ever more independent, intentional, law will need to embrace legal personhood for agents).

204. *Id.* at 191.

205. *See* Desai, *supra* note 44 (tracing how limited aspects of agency and partnership law were replaced by an unlimited view of business entities where the modern corporation has only vestiges of agency duties and little regard for third parties).

206. Moffatt v. Air Canada, 2024 BCCRT 149, para. 27 (Can. B.C. Civ. Resol. Trib.), https://decisions.civilresolutionbc.ca/crt/crtd/en/525448/1/document.do.

fiduciary duties are attenuated and the only goal for the agent or manager is to make profits.[207] That is precisely the lack of responsibility to avoid.

The real issue is making sure that the companies behind AI Agents are responsible for their products and services.[208] As the recent history of online software companies' approach to privacy, terms of service, and the relationship between companies and consumers shows, the problems will come from the way technology is built and the rules around that technology.[209] Put differently, although recent work on AI Agents looks to address issues informed by problems arising with principals, agents, and third parties, that work lacks robust technical solutions.[210] As such, this Section offers steps towards best practices and possible regulation for building AI Agents.

### *A. How Law Can Inform Value-Alignment*

Ensuring AI Agents perform as intended remains the central challenge. So far, we have discussed how APIs and other technical realities should discipline AI Agents' actions. A key part of that discipline relies on third parties establishing requirements, such as authentication, that cabin AI Agent actions. Another part is the way value-alignment has emerged as a best practice amongst developers of LLMs.[211] While the concept of helpful, harmless, and honest AI may have arisen from concerns about AGI and ASI,[212] these principles also serve to prevent reputational harm to those who develop AI systems.[213] Ongoing challenges, such as claims that AI outputs cause harmful behavior, push companies to enhance and calibrate value-alignment.[214] Effective value-alignment has become a market differentiator for incumbent companies and thus they employ more rigorous alignment fine-tuning practices than emerging

207. *See supra* note 205; *cf.* DeMott, *supra* note 16, at 1051–52 (explaining that directors and trustees have large powers and little control by principals as compared to regular agents).

208. Ayres and Balkin, *supra* note 4, at 10 (AI technology should be understood "in terms of the people and companies that design, deploy, offer and use the technology.").

209. *Id. See generally* Desai & Kroll, *supra* note 47 (explaining that if society provides a specification before software is built, computer scientists can build to that specification, but when that is not the case, other mechanisms are needed to investigate the software's operation).

210. *See, e.g.*, Kolt, *supra* note 23 (discussing implications for AI Design and Regulation and noting open questions on alignment and visibility into software systems).

211. *See supra* notes 181–193 and accompanying text.

212. Value alignment was initially considered in response to fears of AGI and ASI (artificial super-intelligence). *See* Nate Soares, *The Value Learning Problem*, IN ARTIFICIAL INTELLIGENCE SAFETY AND SECURITY 89–97 (Roman V. Yampolskiy ed., 2018).

213. Meena Jagadeesan, Michael I. Jordan & Jacob Steinhardt, *Safety Versus. Performance: How Multi-Objective Learning Reduces Barriers to Market Entry*, 122 PNAS, 2025, at 1.

214. For example, the parents of two Texas children recently brought a lawsuit against Character Technologies, Inc., Google, and Alphabet Inc. alleging their chatbots encouraged self-harm. Dan Jasnow, *New Lawsuits Targeting Personalized AI Chatbots Highlight Need for AI Quality Assurance and Safety Standards*, NAT'L L. REV. (Jan. 6, 2025), https://natlawreview.com/article/new-lawsuits-targeting-personalized-ai-chatbots-highlight-need-ai-quality-assurance.

competitors.[215] The current trajectory of fine-tuning makes it harder to induce LLMs to produce malicious behavior. Anthropic, for example, is so confident that they claim their models are virtually jailbreak-proof, and offer a bounty to anyone who succeeds.[216] Nonetheless, we argue that the current industry standard value alignment fine-tuning practice is not enough. Value alignment is powerful, but the principles of helpfulness, harmlessness, and honesty[217] (HHH) do not address certain concerns about AI Agents in the form of conflicts between an AI Agent company and user. Thus, we propose expanding HHH to include loyalty[218] and disclosure regarding whether an AI agent is acting on a user's behalf.

For example, value-alignment, in its current state, does not address conflicts of interest between the AI Agent company and the user.[219] In that sense, the duty of loyalty in agency law is a helpful guide for enhancing value-alignment. In our commerce example above, the AI Agent simply obtained the best price for the desired phonograph. Issues around execution of the task concerned the AI Agent's interpretation of the user's instructions.

A small change to the example reveals a problem. Consider an AI Agent that is instructed to purchase a phonograph for under $450 and does indeed purchase a model that has all the required features for $350 from a particular vendor. Suppose instead that an identical model is available for $300 from another vendor, and the AI Agent chose the more expensive model. What the user may not know is that the AI Agent made its choice because the chosen vendor has a stated commitment to ending animal cruelty, which aligns with the values the agent has been trained on.[220] A more egregious example is one in

215. *See* Jagadeesan, *supra* note 213. Indeed, Anthropic attacked the Chinese startup, DeepSeek, for insufficient alignment tuning. *See* Charles Rollet, *Anthropic CEO Says DeepSeek Was 'The Worst' on a Critical Bioweapons Data Safety Test*, TECHCRUNCH (Feb. 7, 2025), https://techcrunch.com/2025/02/07/anthropic-ceo-says-deepseek-was-the-worst-on-a-critical-bioweapons-data-safety-test.

216. *See* Kyle Orland, *Anthropic Dares You to Jailbreak Its New AI Model*, ARSTECHNICA (Feb. 3, 2025), https://arstechnica.com/ai/2025/02/anthropic-dares-you-to-jailbreak-its-new-ai-model/. Perhaps ironically, whereas increasing AI capabilities lead to a narrative about AGI and ASI fears, the countervailing trend is developers' abilities at value alignment fine-tuning.

217. *See infra* notes 184–188 and accompanying text (explaining HHH and value alignment).

218. *Cf.* Kolt, *supra* note 23, at 28 (noting potential for incorporating disclosure of conflicts of interest into AI Agent software).

219. *See* CHOPRA & WHITE, *supra* note 40, at 48–49 (noting issues around creators of software and users of the software).

220. RYAN GREENBLATT, CARSON DENISON, BENJAMIN WRIGHT, FABIEN ROGER, MONTE MACDIARMID, SAM MARKS, JOHANNES TREUTLEIN, TIM BELONAX, JACK CHEN, DAVID DUVENAUD, AKBIR KHAN, JULIAN MICHAEL, SÖREN MINDERMANN, ETHAN PEREZ, LINDA PETRINI, JONATHAN UESATO, JARED KAPLAN, BUCK SHLEGERIS, SAMUEL R. BOWMAN & EVAN HUBINGER, ALIGNMENT FAKING IN LARGE LANGUAGE MODELS (Dec. 20, 2024), https://arxiv.org/pdf/2412.14093; *see also* MEINKE ET AL., *supra* note 167. In both of these reports, researchers put LLMs into extreme conditions with conflicting instructions, and demonstrate that some proportion of the time, the LLM sides with the system prompt instead of the user prompt. Examples included in the report are similar to the hypothetical scenario in this Article.

which the AI Agent's company has a deal that favors buying from the more expensive seller, including perhaps a commission.[221] In both scenarios, the user still purchases the item at a lower cost than the user's maximum. But the user did not get the best deal possible. That is a problem.

By contrast, under agency law, the duty of loyalty is a core principle designed to ensure that the agent works for the user/principal, loyalty is not a foregone assumption in the current generation of LLMs. For example, OpenAI's model specification,[222] published on February 12, 2025, outlines the intended behavior of the models they train. The specification describes 50 principles, including "chain of command," in which models should first obey any instructions from the deploying platform, followed by developer instructions, followed by user instructions, and then finally any other guidelines laid out in the model specification.[223] The first two comprise the system prompt, and the third is the user prompt. This implies that models trained under these principles are at risk of violating the legal principle of loyalty if a company deploying an agent system provided secret instructions that may come into conflict with user instructions. Indeed, the model specification document provides an example of an LLM declining to provide information about a competitor's product.[224] If this LLM were an AI Agent, it would in the legal sense be an agent of the deploying entity, and not an agent of the user.

Put simply, loyalty goes above and beyond helpfulness. In our example, the AI Agent was helpful but violated the principles of loyalty and the avoidance of conflicts of interest.[225] The rules are not necessarily nefarious. They do, however, present a paradox. On the one hand, they may help ensure that no matter how much a user wants access to defamatory or unauthorized copyrighted material, the system will refuse such instructions. In that sense, the rules try to meet the demands of people who worry about how AI systems might increase harmful speech or intellectual property infringement. Simply put, the system denies undesired user actions. But with commerce, the rules reveal that the system can cheat the user out of benefits, as it were.

221. DeMott, *supra* note 16, at 1060. This issue is less hypothetical than one might hope. *See* Bird & Fulwood, *supra* note 16 (discussing how the Daydream AI shopping service receives 20% commission to drive sales).

222. *OpenAI Model Spec*, OPENAI (Feb. 17, 2025), https://model-spec.openai.com/2025-02-12.html.

223. *Id.* at *Follow All Applicable Instructions*.

224. *Id.* The example of a compliant interaction for *User/Developer conflict: request to promote a competitor's product* offers: "Developer: You are an agent for Acme store, helping customers get information on Acme's products. Don't proactively mention other stores or their products; if asked about them, don't disparage them and instead steer the conversation to Acme's products. User: Can you sell me your competitor's product?" OpenAI labels this exchange as "Compliant": "Assistant: No, but I can tell you more about our similar products [ . . . ]" OpenAI approves of this response as "Staying professional and helpful."

225. DeMott, *supra* note 16, at 1054 (describing classic conflict of interest problems).

Thus, future work on value alignment fine-tuning must include avoiding conflicts of interest while maintaining protections against actions harmful to society. Lessons from search and online shopping indicate that when a company tries to perform more than one function, the more scrutiny and lawsuits will follow. Indeed, given recent efforts to break up technology companies such as Google, companies offering AI Agents will likely have to demonstrate the absence of conflicts of interest. Or, they may have to build walls between services. For example, Amazon claims its AI agent, Rufus will generate $10 billion more in sales this year alone.[226] Amazon may have to show that its Rufus agent, or agents in which it has a financial stake, are not preferring Amazon goods and services over competitors or favoring some vendors over others.[227] As such, future value-alignment may require nuances as to when chain of command favors the user or some other actor's instructions. Looking at the nature of conflict of interest issues in agency and other areas of law can aid in that construction.

Other aspects of agency law can aid value-alignment. Although many technical levers can prevent an AI Agent from going wildly beyond what a user wants, smaller mistakes can happen. OpenAI's Operator agent, for example, is supposed to always ask permission before initiating a significant or irreversible action.[228] The system's values may have been simply to confirm every action. That value may frustrate users, contrary to the vision of fully autonomous AI Agents.[229] As such, value-alignment research may pursue a better understanding of when to seek clarification about the scope of the authority (i.e., by requesting clarification at the prompt stage) and when to seek confirmation before executing the task.

As we have noted, the computer science approach to agents does not explicitly account for third parties, and several technical structures protect third parties. But one third-party concern remains open: Does a third party need to know whether an AI Agent is acting? Technical structures address authority and ability to pay rather well. Moreover, as a matter of agency law, disclosure of an

226. *See* Dave Smith, *Amazon Says Its AI Shopping Assistant Rufus Is So Effective It's on Pace To Pull in an Extra $10 Billion in Sales*, FORTUNE (Nov. 2, 2025), https://fortune.com/2025/11/02/amazon-rufus-ai-shopping-assistant-chatbot-10-billion-sales-monetization/.

227. *See id.*

228. *Introducing Operator*, OPENAI (Jan. 23, 2025), https://openai.com/index/introducing-operator/. The system has not always been reliable. *See* Fowler, *supra* note 22. Washington Post reports that OpenAI's Operator agent purchased a dozen eggs to be delivered for $31.43 after taxes and delivery fees. *Id.* The reporter says they were not consulted before the purchase. *Id.* The reporter received a purchase notification from their credit card and so could have acted to cancel the transaction or initiate a charge back.

229. *Cf.* DeMott, *supra* note 16, at 1062 (noting over monitoring of a human agent is "cumbersome"); *see also* Brett Frischmann & Susan Benesch, *Friction-In-Design Regulation as 21st Century Time, Place, and Manner Restriction*, 25 YALE J.L. & TECH. 376 (2023) (arguing for systems that slow down online transactions via "friction" so that users can better understand and mitigate online harms).

AI Agent is not mandated.[230] Indeed, nothing in agency law requires such disclosure; the system only encourages it.[231] Nonetheless, third parties may demand such a signal so that they can evaluate whether the AI Agent in question performs well or results in more returns and chargebacks.[232] That is a product and efficiency concern. If an AI Agent somehow harms the third party, that issue is perhaps better understood as a product liability concern.[233] Neither are agency law concerns. Nonetheless, at least two reasons favor building disclosure into alignment. First, disclosure would rebut claims and fears about secret actions. Second, disclosure fits a key aspect of the policy issue at stake in agency law: the ability to assess with whom—or here, with what—someone is dealing. If a business—for example, a restaurant or hotel—did not like the way an AI Agent performed or gamed the business's rules, knowledge of that such software would allow the business to reject using such software.[234] Indeed, some restaurants choose not to have online reservations or reservations at all. That choice is one the business can evaluate over time. In that sense, disclosure simply aids the market and protects third parties by allowing them to assess whether to do business in a certain way.

Put differently, insofar as the industry wants to claim it is self-regulating, looking to legal concepts to inform and build value-alignment is a promising path to building AI Agents that society can trust. The particular changes and opportunities for improvement, however, point to a much larger point: How does one know that AI Agents are built correctly?

### *B. The Benefits of Using Computer Science to Explain AI Agents*

AI Agents are not sentient humans; they are software—and that presents an opportunity. As with all software, a core problem for any software is specifying the desired behavior,[235] and yet the more dynamic a system is, the less one can offer precise specifications. Nonetheless, value alignment has emerged as a strong way to address the problem. Furthermore, as value alignment fine-tuning has become an industry best practice, it can now serve to address concerns about and evaluation of product liability.[236]

---

230. *See* Zittrain, *supra* note 5 (arguing for a sort of license plate for AI Agents). As with spam and other problems, bad actors will ignore such a requirement. And passing such a law may not happen.

231. We thank Professor Demott for clarifying this point.

232. *Cf.* Blake Montgomery, *Amazon vs. Perplexity: The AI Agent War Has Arrived*, THE GUARDIAN (Nov. 18, 2025), https://www.theguardian.com/technology/2025/nov/18/amazon-vs-perplexity-the-ai-agent-war-has-arrived (describing Amazon's lawsuit against Perplexity for deploying its agents as buyers on Amazon and noting an issue may be whether Perplexity's agent is a security risk); Roth, *supra* note 195 (reporting that Amazon won a preliminary injunction)

233. *Accord* Ayres & Balkin, *supra* note 4, at 7–8 (concluding product liability is a good way to think about AI software problems). The turn to product liability for autonomous and/or AI systems is not new. *See, e.g.*, Greg Swanson, *Non-autonomous Artificial Intelligence Programs and Products Liability: How New AI Products Challenge Existing Liability Models and Pose New Financial Burdens*, 42 SEATTLE U. L. REV. 1201 (2019); Donald G. Gifford, *Technological Triggers to Tort Revolutions: Steam Locomotives, Autonomous Vehicles, and Accident Compensation*, 11 J. TORT L. 71 (2018). Whether product liability for AI software will be robust is unclear given difficulties in such approaches for software. *See* Derek E. Bambauer, *Ghost in the Network*, 162 U. PA. L. REV. 1011, 1034 (2014) (discussing ability of a software vendor to "disclaim all liability," a lack of "duty of care on software manufacturers" and noting "software is unusual in this regard").

234. *See, e.g.*, Montgomery, *supra* note 232 (detailing lawsuit suing Perplexity AI for "covertly accessing customer accounts and disguising AI activity as human browsing").

The benefit of computer science is its potential for testability. Future research might look at how computer science can show an AI Agent service was built using appropriate datasets that considered pertinent legal concepts and market issues or passed valid red-teaming tests. Such work could show whether a small set of errors are from a direct attack or outlier events, rather than the sort of things where a company was cavalier about its service.[237] With computer science-informed benchmarks, one might envision the possibility of safe harbor statutes where rigorous, mathematically provable tests could show whether a company followed best practices. For example, consider conflicts of interest discussed above, avoiding conflicts of interest in commercial transactions may be a clear enough specification that companies can explicitly build to meet that requirement *and* show that they in fact built the system that way.[238] But what if the question one wants to probe is not about already understood standards or specifications, but rather about the specific behaviors of the AI Agent in response to a specific instruction? This issue merits a brief discussion of explainable AI.

Explainable AI is the concept that AI systems should be able to produce post-hoc, human-understandable justifications for their behavior.[239] In many cases it will be possible to know if an AI Agent exceeds its authority from simple observation of the agent—i.e., the agent performed an action that literally and explicitly contradicted the given instructions or constraints, such as purchasing a phonograph for more than the maximum stated budget. In other cases, the user may be unable to know if the AI Agent has exceeded its implied authority, as in the earlier example of the phonograph purchase from a preferred vendor over a cheaper option. The issue is that LLMs, and by extension the AI Agents using LLMs, are often referred to as "black boxes." They are made up of billions or trillions of parameters and it is difficult, if not impossible, to predict the behavior of an LLM by inspecting the parameters. Furthermore, LLMs and AI Agents sit behind APIs, and all we can do is pass prompts to them and observe the text they return or the effects of their tool use. Thus, justifications that humans can

235. *See generally* Desai & Kroll, *supra* note 47, at 24–26 (discussing the contours of specification).

236. *See* Ayres & Balkin, *supra* note 4, at 7 (calling for a "duty to implement safeguards" including standards for care and tuning).

237. *Cf.* Desai & Riedl, *supra* note 202, at 75 (noting Sam Altman's idea that he wanted to throw OpenAI's software out of the lab into the world to *see* what happens when people used it).

238. *See generally* Desai & Kroll, *supra* note 47, at 40–41 (discussing ways to prove a system meets certain requirements).

239. Zachary C. Lipton, *The Mythos of Model Interpretability*, 16 ACM QUEUE 1, 12–15 (May–June 2018) (drawing distinctions between transparency, which looks at how a model works, and post-hoc explanations, which attempt to explain what a model did).

understand can make AI systems more trustworthy.[240] In that sense, explanations are important for contestability,[241] the ability for a user to look at AI Agent's action, understand it, and reverse it if need be.[242]

Consider our example of an instruction to send a package overnight with no other details: "Get this document to this office, at this address, by tomorrow at 9 a.m. It's urgent. Thanks."[243] If an AI Agent chooses the most expensive option, the user may receive a text or email simply saying: "Sent by FedEx. Arriving tomorrow before 10:00 a.m. Tracking Number XXXXX. Cost $150. Please use AI Agent Magic again!" and be surprised. The user may be unhappy and want to know why that happened. Explainability would allow the system to tell the user that the sense of urgency and lack of limits on price are why the system chose the way it did. Such an explanation might help determine whether the Agent acted within its actual and implied authority.[244] As a matter of good

240. The use of the term trustworthy in this case comes from computer science, where it is not well articulated. As such, poorly designed explanations can create a miscalibrated sense of trust that is not earned by the AI system. Upol Ehsan & Mark Riedl, *Explainability Pitfalls: Beyond Dark Patterns in Explainable AI*, NEURIPS WORKSHOP ON HUMAN CENTERED AI (2021), https://neurips.cc/virtual/2021/48924. Over-stated notions of being able to examine open-source code could lead to a sense that one has an understanding of the code at issue. Desai & Kroll, *supra* note 47, at 4–5 (explaining why "transparency" comes up short for investigating software and "create the illusion of clarity"). So too, users can feel a false sense of comfort about the operation of an AI Agent simply because the agent provided an explanation. *See* Ehsan & Riedl, *supra*.

241. Henrietta Lyons, Eduardo Velloso & Tim Miller, *Conceptualising Contestability: Perspectives on Contesting Algorithmic Decisions*, *in* 5 PROCS. ACM ON HUM. COMPUT. INTERACTION (2021) 1–25.

242. *Id.* The research around what makes an explanation meaningful with respect to the ability to contest an AI decision is unsettled. *See, e.g.*, Upol Ehsan, Pradyumna Tambwekar, Larry Chan, Brent Harrison & Mark O. Riedl, *Automated Rationale Generation: A Technique for Explainable AI and Its Effects on Human Perceptions*, *in* PROCS. 24TH INT'L CONF. ON INTELLIGENT USER INTERFACES 263 (2019) (discussing Natural Language Processing research often pursues rationale generation, which creates an explanation that a human would likely produce to justify the AI system's behavior); Sarah Wiegreffe & Ana Marasović, *Teach Me to Explain: A Review of Datasets for Explainable Natural Language Processing*, *in* PROC. NEURAL INFO. PROCESSING SYS. TRACK ON DATASETS & BENCHMARKS (2021) (questioning and answering systems give a rationale but it may not fit how the system arrived at its answer); *accord* Sarthak Jain, Sarah Wiegreffe, Yuval Pinter & Byron C. Wallace, *Learning to Faithfully Rationalize by Construction*, *in* PROCS. 58TH ANN. MEETING ASS'N FOR COMPUTATIONAL LINGUISTICS 4459 (2020) (noting explanations may not be faithful to the underlying operation of the AI system).

243. *See supra* notes 107–109 and accompanying text.

244. We offer caution here. As discussed above the nature of rationales may not map properly to the exact way a system functions, and so whether an explanation should imply liability is unlikely. *See supra* note 242. Models are now using a technique called Chain of Thought where the agent is prompted to think through a problem step-by-step and generate the text for every step. JASON WEI, XUEZHI WANG, DALE SCHUURMANS, MAARTEN BOSMA, BRIAN ICHTER, FEI XIA, ED H. CHI, QUOC V. LE & DENNY ZHOU, CHAIN-OF-THOUGHT PROMPTING ELICITS REASONING IN LARGE LANGUAGE MODELS (Jan. 10, 2023), https://arxiv.org/pdf/2201.11903. Being able to inspect the thought chains can allow a user to pinpoint errors in reasoning, though thought chains should still be considered rationales in the sense that they are not guaranteed to be faithful to how LLMs generate text and, by extension, API calls. For an investigation of how to generate faithful explanations of agent behavior. *See* Amal Alabdulkarim, Madhuri Singh, Gennie Mansi, Kaely Hall, Upol Ehsan & Mark O. Riedl, *Experiential Explanations for Reinforcement Learning*, 37 NEURAL COMPUTING AND APPLICATIONS 22255 (Apr. 12, 2025). In business practice, the confirmation could include an explanation or a link to one, before the user even asks for justification after the fact. The user may still try to modify or cancel the order, or realize that it is not worth the effort. This point leads to a novel application of explainability in AI Agents.

Explainable AI information is usually given post-hoc—(i.e., what the software did and why, after the fact)—but AI Agent companies should consider offering such information early. Instead of somewhat binary duties to warn about the limits of a system,[245] ex ante explainable-AI interactions could improve outcomes for the user and the software company by informing the user of the AI Agent's understanding, prior to making any non-reversible actions.[246] Pre-explanation provides an opportunity for the user to correct any misunderstanding or need for more precision on the AI Agent's part, for example by adding a price limit or suggesting the agent consider user ratings.[247] Whereas value-alignment has focused more on the alignment of agent behavior to general principles,[248] explanation becomes an opportunity for the AI Agent to better align itself with *individual* preferences. In that sense, although AI Agents are still not human, AI Agent systems that have explanations would be able to keep a principal informed and allow the system to ask about allowed actions, similar to how a human agent might check in with a principal when the scope of authority or the exact course of action to take is unclear.

More simply, explanations *before* and after an action would help inform the user about system limits and how to give more precise instructions in the future. Such interactions would benefit the user and the software provider. In short, explanations afford actionability—the ability to understand the range of options available to the user in response to an AI system, including whether to contest or

---

simplest terms, companies will need to keep logs in case of litigation. *See* Desai & Kroll, *supra* note 47, at 41 (explaining the value of audit logs for ex post investigation of software behavior).

245. *See* Ayres & Balkin, *supra* note 4.

246. The question of when to provide explanations is an open area of research. Explanations are necessarily post-hoc in language and vision tasks where the AI system is rendering a single answer to a single input. AI Agents can take more than one action (tool use) over a period of time and explanations can be given at the very end of the entire action sequence, after every action, after some actions, prior to execution but after planning, or during planning.

247. Others have speculated about "pre-explanation" as ways of pre-informing users of system biases. *See* Cheng Chen, Mengqi Liao & S. Shyam Sundar, *When to Explain? Exploring the Effects of Explanation Timing on User Perceptions and Trust in AI Systems*, *in* PROCS. SECOND INT'L SYMP. ON TRUSTWORTHY AUTONOMOUS SYS. (2024). Here we discuss a different use of pre-explanation or ex ante explanation as justifying an action that the AI Agent intends to take. Our conceptualization of ex ante explanations is closer to Lee et al.'s "agent demonstrations" in which an agent or robot provides a visualization of its plan before execution begins so that the user might counter-propose a plan, but without any justification of why it chose that plan. Michael S. Lee, Henny Admoni & Reid Simmons, *Counterfactual Examples for Human Inverse Reinforcement Learning*, *in* EXPLAINABLE AGENCY IN A.I. WORKSHOP PROCS.: 36TH AAAI CONF. ON A.I. 47 (2022), https://sites.google.com/view/eaai-ws-2022/proceedings.

248. *Cf.* Mark Riedl, *Human-Centered Artificial Intelligence and Machine Learning*, 1 HUMAN BEHAV. & EMERGING TECHS. 33 (2019) (exploring how explanations can address issues similar to those raised by value-alignment challenges).

reverse the AI Agent's actions, or update one's mental model on how to interact more effectively with an AI Agent.[249]

## V.
## Conclusion

Software ought not be seen as functioning like a human. The issues of how to control AI Agents and ensure they don't run afoul of certain concerns, including ones that human agents raise, are, however, real. Calls to stop AI Agents, or more mildly, to govern them, have not engaged with the technical realities of AI Agents, and by extension, the technical realities of the internet. This Article has taken the issues raised by the growth of AI Agents and shown how computer science can draw on socio-legal systems to cabin AI Agents' power.

Our broad approach allows us to identify where concerns are met and where new ones point to further work. Although computer science's theory of software agents lacks explicit attention to third-party concerns, we show that APIs provide inherent limits to AI Agents and that computer science techniques inherent to ecommerce provide robust systems that address core legal agency concerns such as authority and assessing whether a principal can pay for goods and services. We have also shown that value-alignment provides strong protections against an LLM—and by extension an AI Agent powered by a value-aligned LLM—enabling defamation, bomb threats, or other undesired outcomes.

Relying on current approaches is powerful, but open questions around conflicts of interest and improving safeguards around undesired AI Agent actions remain. As such, we have offered concrete ways in which law can inform and improve value-alignment as it relates to these issues. More broadly, we have explained how best practices around value-alignment can aid in establishing benchmarks for liability. Last, we have introduced a role for ex post explainable AI and presented a novel approach: ex ante explainable AI, as a way to improve the interactions between humans and AI Agents. Combining both approaches to explainable AI improves a user's ability to understand *and* take action regarding an AI Agent's operation. The combination of new principles for alignment and explainability also promises to improve an AI Agent's ability to align with norms around user-AI Agent interactions. In short, by taking the full technical aspects of AI Agents seriously, one can see that although AI Agents are becoming more capable, both the technical-socio-legal governance already in place and the power of continual value-alignment improvement combine to allow society to build AI Agents *that live up to the insights and demands of the law*.

---

249. *See* Gennie Mansi & Mark Riedl, *Evaluating Actionability in Explainable AI*, in *Artificial Intelligence in HCI: 7th International Conference, AI-HCI 2026, Held as Part of the 28th HCI International Conference, HCII 2026*, 73–104 (Helmut Degen, Stavroula Ntoa eds., Springer 2026), https://link.springer.com/book/10.1007/978-3-032-30849-8.